\documentclass[aps,pre,reprint,amsmath,amssymb,longbibliography]{revtex4-2}

\usepackage{graphicx}
\usepackage{dcolumn}
\usepackage{booktabs}
\usepackage{bm}
\usepackage{hyperref}
\usepackage{xcolor}
\newcommand{\rev}[1]{#1}
\hypersetup{hidelinks}

\begin{document}

\title{Degenerate four-wave mixing in a \texorpdfstring{$\mathcal{CPT}$}{CPT}-symmetric coupler with intermodal dispersion}
\date{\today}

\author{Nguyen Duc Anh Quan}
\affiliation{School of Materials Science and Engineering (SMSE), Hanoi University of Science and Technology (HUST), No. 1 Dai Co Viet Street, Hanoi, Vietnam}

\author{Do Duc Tho}
\affiliation{School of Engineering Physics (SEP), Hanoi University of Science and Technology (HUST), No. 1 Dai Co Viet Street, Hanoi, Vietnam}
\author{Marek Trippenbach}
\affiliation{Faculty of Physics, University of Warsaw, Pasteura 5, Warsaw, 02-093, Poland}
\author{Nguyen Thi Dung}
\affiliation{Faculty of Natural Sciences, Hong Duc University, 565 - Quang Trung Street, Thanh Hoa, Vietnam}
\author{Tran Thi Hai}
\affiliation{Faculty of Natural Sciences, Hong Duc University, 565 - Quang Trung Street, Thanh Hoa, Vietnam}
\author{Nguyen Viet Hung}
\email{hung.nguyenviet1@hust.edu.vn}
\affiliation{School of Materials Science and Engineering (SMSE), Hanoi University of Science and Technology (HUST), No. 1 Dai Co Viet Street, Hanoi, Vietnam}

\date{\today}

\begin{abstract}
\rev{Four-wave mixing provides a simple setting in which dispersion, nonlinearity, and non-Hermiticity compete to select resonant energy-transfer channels. We study degenerate four-wave mixing in a Kerr dual-core coupler with balanced gain and loss and frequency-dependent intercore coupling. The dispersive coupling changes the symmetry from conventional $\mathcal{PT}$ symmetry to a combined $\mathcal{CPT}$ symmetry and reshapes the two-branch linear spectrum. We determine the unbroken-$\mathcal{CPT}$ domain and classify the branch configurations that can satisfy the degenerate phase-matching condition. In the parameter ranges examined, three resonant channels persist over broad regions, whereas a same-branch channel appears only close to the symmetry-breaking threshold. In this near-threshold regime, a single pump can simultaneously satisfy two distinct nonzero sideband resonances. Direct pulse simulations confirm the predicted resonances and reveal secondary-wave generation and multifrequency cascades near eigenmode coalescence. A reduced three-wave model captures the initial dynamics away from the exceptional point but loses accuracy as the modal basis becomes ill-conditioned. These results show how dispersive coupling reorganizes resonances, group-velocity mismatch, and nonlinear energy exchange in a non-Hermitian wave system, and they identify the exceptional-point region as a regime where a few-mode description can break down.}
\end{abstract}

\maketitle

\section{Introduction}
\rev{Resonant energy transfer among weakly nonlinear waves is controlled not only by the strength of the nonlinearity but also by the geometry of the underlying linear spectrum. Degenerate four-wave mixing (FWM) is a particularly transparent example: two pump quanta are converted into a signal--idler pair while both frequency and propagation-constant matching are satisfied. In optical systems the same third-order interaction underlies parametric amplification, phase conjugation, wavelength translation, frequency-comb formation, and correlated-pair generation \cite{carman1966,agrawal2019,boyd2020}. For a single dispersion branch, the matching problem is largely set by material and waveguide dispersion. Once several modes or supermodes participate, however, the branch index and the internal modal composition become additional dynamical variables, so the set of allowed resonances can change qualitatively.}

\rev{This extra modal freedom has already been exploited in few-mode fibers and integrated waveguides. Intermodal phase conjugation and Bragg scattering can connect spectral intervals that are difficult to reach through intramodal mixing, and mode-selective conversion has been demonstrated in both fiber and silicon platforms \cite{friis2016,signorini2018,lacava2019}. At the same time, coherent buildup is sensitive to how the linear modes are mixed: random linear mode coupling can lower the conversion and reduce the useful parameter interval \cite{xiao2014}. These observations are important for the present problem because they show that modal dispersion is not merely a small correction to a nonlinear interaction. It can reorganize the resonance conditions themselves and thereby alter which energy-transfer pathways are dynamically accessible.}

\rev{A dual-core coupler is the minimal setting in which this branch structure can be studied without introducing a large number of spatial modes. Even when each isolated core is single mode, the symmetric and antisymmetric supermodes generally have different group delays \cite{chiang1995}. In the core representation the same physics appears as a coupling coefficient that depends on frequency. Its first- and second-order frequency dependence can produce temporal walk-off, pulse broadening, breakup, and marked changes in nonlinear switching \cite{chiang1997,paiva1999,shum2002,liu2003}; dispersive coupling can also stabilize spatiotemporal states that would otherwise be unstable \cite{kartashov2015}. Consequently, resonant wave mixing in such a system is governed by two intertwined dispersive mechanisms: the ordinary propagation dispersion of each arm and the dispersion of the intercore coupling.}

\rev{Balanced gain and loss add a second ingredient that is especially relevant from the viewpoint of nonlinear dynamics: the linear operator becomes non-Hermitian while its spectrum may remain entirely real below a symmetry-breaking threshold. In that regime the eigenmodes are nonorthogonal, and at an exceptional point both their eigenvalues and eigenvectors coalesce \cite{ruter2010,konotop2016}. Nonlinear extensions of such systems support continuous soliton families, asymmetric transport, and strongly phase-sensitive evolution \cite{alexeeva2012,ramezani2010}. Four-wave-mixing schemes with related non-Hermitian symmetries have also been studied in cold and thermal atomic vapors \cite{jiang2019,niu2024}, while nonreciprocal FWM has been proposed as a mechanism for optical isolation \cite{munoz2022}. These examples motivate a broader question: how does a non-Hermitian spectral degeneracy modify the resonance structure and the validity of reduced nonlinear-wave descriptions?}

\rev{An important reference point is the work of Wasak et al. on resonant FWM in a Kerr $\mathcal{PT}$ coupler with frequency-independent coupling \cite{wasak2015}. In that case balanced gain and loss shift the matching between the slow and fast branches, open a channel that is absent in the conservative limit, and allow a secondary resonance to generate a fifth wave. Because the coupling is independent of frequency, however, the internal modal phase is the same for all participating frequencies and a common carrier-frequency shift can be removed by a Galilean transformation. The dynamics considered here is more general: once the coupling itself is dispersive, both the branch separation and the eigenvector composition vary across the interacting spectrum.}

\rev{Coupling dispersion also changes the appropriate symmetry operation. The first-order derivative in the coupling changes sign under temporal inversion, and the equations become invariant only when temporal parity and time reversal are supplemented by an exchange of the two cores. We denote this core exchange by $\mathcal C$, so that the relevant operation is $\mathcal{CPT}$; here $\mathcal C$ is a geometrical exchange of the two waveguide arms rather than charge conjugation. Zezyulin et al. used this symmetry to derive the real-spectrum condition and to construct stable vector solitons in the nonresonant regime \cite{zezyulin2017}. The present problem probes a complementary regime. Terms that average out in a nonresonant description become the slowly varying resonant terms when the FWM matching conditions are satisfied, and their strength now depends explicitly on frequency through the non-Hermitian eigenvectors.}

\rev{The classification of resonant branches is closely related to the treatment of FWM in spin--orbit-coupled Bose--Einstein condensates \cite{hung2020}, where the sideband roots and group velocities separate dynamically distinct channels. This analogy is useful because it emphasizes features that are not specific to a particular optical device. The present coupler nevertheless introduces two additional difficulties: the linear modes form a nonorthogonal basis, and their internal phase varies with frequency because of the dispersive coupling. As a result, phase matching, pulse overlap, and nonlinear overlap coefficients become mutually dependent rather than separable ingredients.}

\rev{We therefore organize the study around three questions. First, which branch triples can satisfy degenerate matching while the $\mathcal{CPT}$ spectrum remains real? Second, under what conditions can a single pump support more than one nonzero sideband pair, so that several resonant channels coexist? Third, how far can a low-dimensional three-wave reduction be trusted when the full dynamics involves finite bandwidth, walk-off, gain--loss exchange, and proximity to an exceptional point? We derive the real-spectrum boundary, exclude four branch triples analytically within the weak-dispersion domain, and reduce the remaining phase-matching problem to a cubic equation for the squared sideband separation. Direct pulse simulations are then compared with a biorthogonally projected three-wave model. The comparison reveals not only the resonant conversion channels but also the onset of secondary waves and the loss of accuracy of the reduced description near eigenmode coalescence. Finally, we map the output sideband fraction versus the two coupling-dispersion coefficients to show how the two dispersive orders affect different parts of the nonlinear dynamics.}

\section{Model and \texorpdfstring{$\mathcal{CPT}$}{CPT} symmetry}

\rev{We use the simplest two-field model that retains the three ingredients of interest: Kerr nonlinearity, balanced gain and loss, and frequency-dependent intercore coupling. The two dispersive waveguides are otherwise identical, so changes in the dynamics can be traced directly to the non-Hermitian balance and to the coupling dispersion. Gain is placed in the first arm and an equal loss in the second. In dimensionless variables, the field envelopes obey}
\begin{align}
 i\frac{\partial q_1}{\partial z}
 &= -\frac{\partial^2 q_1}{\partial \tau^2}
 +i\gamma q_1-\mathcal K q_2+\sigma |q_1|^2q_1,
 \label{eq:model-q1}\\
 i\frac{\partial q_2}{\partial z}
 &= -\frac{\partial^2 q_2}{\partial \tau^2}
 -i\gamma q_2-\mathcal K q_1+\sigma |q_2|^2q_2,
 \label{eq:model-q2}
\end{align}
where $q_{1,2}(z,\tau)$ are the slowly varying envelopes, $z$ is the propagation distance, and $\tau$ is retarded time. The coefficient $\gamma>0$ denotes gain in the first arm and equal loss in the second. The Kerr coefficient is normalized to $\sigma=+1$ for self-defocusing and $\sigma=-1$ for self-focusing nonlinearity. The frequency-dependent coupling is represented by
\begin{equation}
 \mathcal K=\kappa_0+i\kappa_1\partial_\tau-\kappa_2\partial_\tau^2,
 \label{eq:coupling-operator}
\end{equation}
where $\kappa_0$, $\kappa_1$, and $\kappa_2$ are the zeroth-, first-, and second-order coupling-dispersion coefficients, respectively. For a Fourier component proportional to $e^{-i\omega\tau}$,
\begin{equation*}
 \widehat K(\omega)=\kappa_0+\kappa_1\omega+\kappa_2\omega^2.
\end{equation*}
\rev{A constant phase shift in one arm changes the sign of $\kappa_0$, while reversal of the temporal coordinate changes the sign of $\kappa_1$. We therefore set $\kappa_0>0$ and $\kappa_1\geq0$. Measuring $z$ in units of $\kappa_0^{-1}$, followed by the corresponding rescaling of $\tau$ and the amplitudes, fixes $\kappa_0=1$. The calculations below use the normalized ordering $0<\kappa_1<1$ and $0<\kappa_2<1$, which represents a weak-to-moderate coupling-dispersion regime after the zeroth-order coupling has been scaled to unity. In physical terms, the first- and second-order frequency-dependent corrections remain smaller than the zeroth-order coupling over the spectral interval sampled by the pulses. This assumption is also useful mathematically because it isolates the physically motivated perturbative regime in which the branch classification can be established analytically. Appendix~A requires only $0<\kappa_2<1$, $\kappa_1^2<4\kappa_2$, and the unbroken-$\mathcal{CPT}$ condition; the stronger numerical bound on $\kappa_1$ is not used in the proof.}

Let $\mathbf q=(q_1,q_2)^T$ and let $\sigma_x$ be the first Pauli matrix. We define temporal parity, time reversal, and core exchange by
\begin{align*}
 \mathcal P &: \tau\mapsto-\tau,\\
 \mathcal T &: (z,i,\mathbf q)\mapsto(-z,-i,\mathbf q^*),\\
 \mathcal C &: \mathbf q\mapsto\sigma_x\mathbf q.
\end{align*}
The combined operation is therefore
\begin{equation}
 (\mathcal{CPT}\mathbf q)(z,\tau)
 =\sigma_x\mathbf q^*(-z,-\tau).
 \label{eq:CPT-transform}
\end{equation}
\rev{Under $\mathcal{PT}$, the operator $i\partial_\tau$ is invariant because complex conjugation and temporal inversion each reverse the sign that appears in this term. The gain--loss part behaves differently: gain in one arm is mapped onto loss and becomes invariant only after the two arms are exchanged. Equations~\eqref{eq:model-q1} and \eqref{eq:model-q2} are therefore invariant under the combined transformation \eqref{eq:CPT-transform}, but not under conventional temporal $\mathcal{PT}$ symmetry alone. Equal Kerr coefficients in the two arms are essential for this invariance. It is useful to distinguish this symmetry of the equations from the spectral phase discussed below: the equations remain $\mathcal{CPT}$ symmetric even when the linear eigenvalues have entered the broken-symmetry regime.}

The total power
\begin{equation*}
 U(z)=\int_{-\infty}^{\infty}\left(|q_1|^2+|q_2|^2\right)d\tau
\end{equation*}
is not generally conserved. Directly from Eqs.~\eqref{eq:model-q1} and \eqref{eq:model-q2},
\begin{equation}
 \frac{dU}{dz}=2\gamma\int_{-\infty}^{\infty}
 \left(|q_1|^2-|q_2|^2\right)d\tau.
 \label{eq:total-power-balance}
\end{equation}
\rev{Equation~\eqref{eq:total-power-balance} makes explicit a point that will matter when interpreting the numerical conversion efficiencies. Balanced material gain and loss do not produce a global conservation law for an arbitrary state; the net power is constant only when the integrated intensities in the two arms remain balanced. The nonlinear waves can therefore exchange power not only among resonant frequencies but also with the gain--loss background, which is why the output quantity used later is described as a spectral power fraction rather than as a conserved photon-conversion efficiency.}

\section{Linear spectrum and eigenmodes}

In the linear limit, we seek plane waves
\begin{equation*}
 \mathbf q(z,\tau)=\mathbf A\,e^{i\beta z-i\omega\tau}.
\end{equation*}
Substitution into Eqs.~\eqref{eq:model-q1} and \eqref{eq:model-q2} gives
\begin{equation*}
 \begin{pmatrix}
 -\omega^2-i\gamma & \widehat K(\omega)\\
 \widehat K(\omega) & -\omega^2+i\gamma
 \end{pmatrix}\mathbf A=\beta\mathbf A,
\end{equation*}
where, after setting $\kappa_0=1$,
\begin{equation*}
 \widehat K(\omega)=1+\kappa_1\omega+\kappa_2\omega^2.
\end{equation*}
The two propagation-constant branches are
\begin{equation}
\begin{aligned}
 \beta_s(\omega)&=-\omega^2+s\varepsilon(\omega),\\
 \varepsilon(\omega)&=\sqrt{\widehat K^2(\omega)-\gamma^2},
 \qquad s=\pm1.
\end{aligned}
 \label{eq:dispersion-branches}
\end{equation}

For $\kappa_2>0$, the minimum of $\widehat K$ occurs at
\begin{equation*}
 \omega_*=-\frac{\kappa_1}{2\kappa_2},
 \qquad
 \widehat K_{\min}=1-\frac{\kappa_1^2}{4\kappa_2}.
\end{equation*}
An all-real spectrum requires $\widehat K_{\min}>0$, or $\kappa_1^2<4\kappa_2$, and
\begin{equation}
 0\leq\gamma\leq\gamma_{\mathcal{CPT}},
 \qquad
 \gamma_{\mathcal{CPT}}=1-\frac{\kappa_1^2}{4\kappa_2}.
 \label{eq:CPT-threshold}
\end{equation}
\rev{Equation~\eqref{eq:CPT-threshold} shows that the real-spectrum window is controlled by the minimum of the frequency-dependent coupling rather than by the gain--loss coefficient alone. If $\kappa_2=0$ while $\kappa_1\neq0$, the linear function $\widehat K(\omega)$ necessarily crosses zero, and no nonzero $\gamma$ can keep the spectrum real for all frequencies. A sufficiently strong positive second-order coupling-dispersion term is therefore not a small quantitative correction: it is the ingredient that prevents the coupling from vanishing at large negative or positive detuning and thereby allows an unbroken $\mathcal{CPT}$ phase over the complete spectral axis \cite{zezyulin2017}. The approach to equality in Eq.~\eqref{eq:CPT-threshold} defines the exceptional-point boundary that will later control the intrabranch FWM channel.}

\begin{figure*}[t]
\centering
\includegraphics[width=0.48\textwidth]{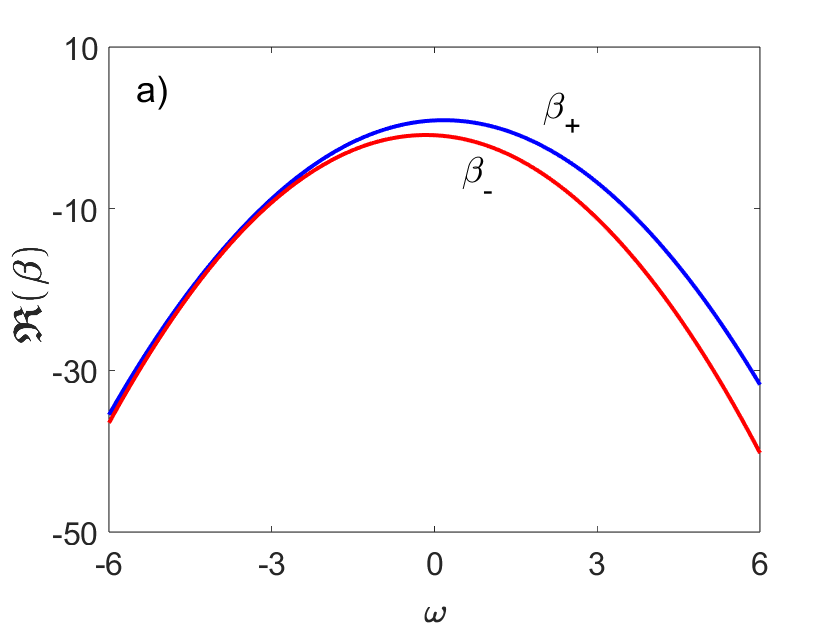}
\includegraphics[width=0.48\textwidth]{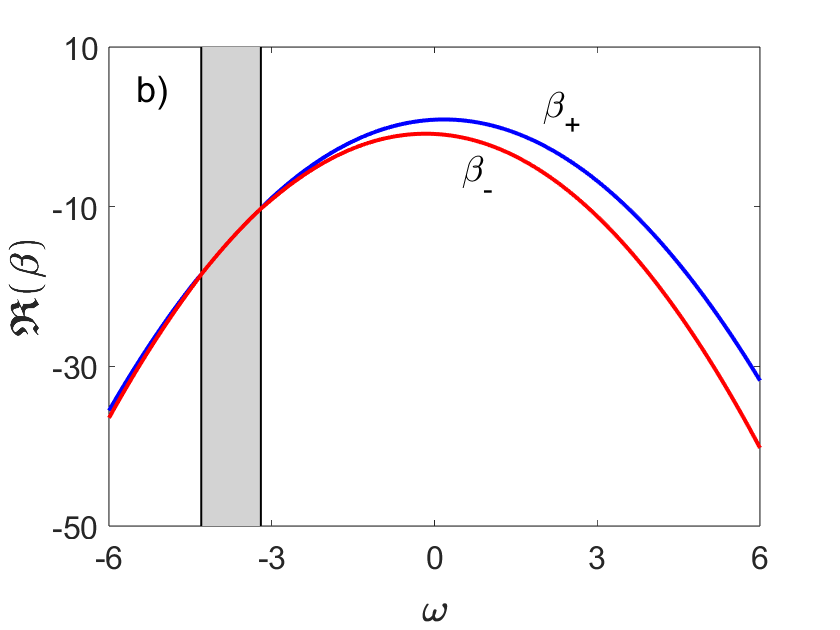}
\caption{Linear dispersion relations for $\{\kappa_0,\kappa_1,\kappa_2\}=\{1,0.3,0.04\}$. (a) Unbroken $\mathcal{CPT}$ phase at $\gamma=0.4$. (b) Broken phase at $\gamma=0.5>\gamma_{\mathcal{CPT}}=0.4375$. The shaded interval marks frequencies for which the branches coalesce and the propagation constants form a complex-conjugate pair.}
\label{fig:linear-dispersion}
\end{figure*}

In the unbroken phase, define the real angle $\phi(\omega)$ by
\begin{equation}
 e^{i\phi(\omega)}=
 \frac{\varepsilon(\omega)+i\gamma}{\widehat K(\omega)},
 \qquad
 \sin\phi(\omega)=\frac{\gamma}{\widehat K(\omega)}.
 \label{eq:internal-phase}
\end{equation}
A convenient normalized right eigenvector is
\begin{equation}
 \mathbf r_s(\omega)=\frac{1}{\sqrt2}
 \begin{pmatrix}
 s e^{-is\phi(\omega)}\\[2pt]1
 \end{pmatrix},
 \label{eq:right-eigenvector}
\end{equation}
so that the complete linear mode is
$\mathbf q_s=\mathbf r_s(\omega)e^{i\beta_s(\omega)z-i\omega\tau}$.
\rev{The eigenvector phase in Eq.~\eqref{eq:right-eigenvector} is frequency dependent, so the modal composition changes across a resonant quartet. This feature is absent in the frequency-independent coupler and is one reason that the nonlinear overlap coefficients below are not constants. In addition, the linear matrix is non-Hermitian, so its right eigenvectors are not orthogonal under the conventional Hermitian inner product. The matrix is nevertheless complex symmetric, $\mathsf H^T(\omega)=\mathsf H(\omega)$. Transposing the right-eigenvalue equation then shows that $\mathbf r_s^T$ is an unnormalized left eigenvector. To project nonlinear terms onto a definite branch we therefore introduce the normalized dual row mode}
\begin{equation}
 \boldsymbol\ell_s^{\dagger}(\omega)=
 \frac{\mathbf r_s^T(\omega)}{\mathbf r_s^T(\omega)\mathbf r_s(\omega)}
 =\frac{\mathbf r_s^T(\omega)}{e^{-is\phi(\omega)}\cos\phi(\omega)},
 \label{eq:left-eigenvector}
\end{equation}
where the dagger labels a left row eigenvector and does not imply $\boldsymbol\ell_s^{\dagger}=\mathbf r_s^{\dagger}$. Directly from Eq.~\eqref{eq:right-eigenvector},
\begin{equation*}
 \mathbf r_s^T\mathbf r_s
 =\frac{1+e^{-2is\phi}}{2}
 =e^{-is\phi}\cos\phi,
 \qquad
 \mathbf r_s^T\mathbf r_{-s}=0,
\end{equation*}
\rev{which gives $\boldsymbol\ell_s^{\dagger}\mathbf r_{s'}=\delta_{ss'}$. This biorthogonal construction is not only a formal replacement for Hermitian orthogonality. It determines how a nonlinear source term is decomposed between the two non-Hermitian branches. The same left--right projection underlies the reduced FWM equations of Ref.~\cite{wasak2015}; here the normalization is frequency dependent because the eigenvectors vary with $\widehat K(\omega)$. The general non-Hermitian spectral setting is reviewed in Ref.~\cite{konotop2016}, and the corresponding frequency-dependent mode transformation for the present $\mathcal{CPT}$ coupler was introduced in Ref.~\cite{zezyulin2017}. Near the exceptional point, $\cos\phi\to0$ and the dual normalization becomes large, foreshadowing the loss of conditioning of the few-mode reduction. We return to this point in Sec.~VI.}

\section{Phase-matching conditions and FWM configurations}

\subsection{General and degenerate matching conditions}

\rev{We next separate the kinematic resonance problem from the subsequent nonlinear dynamics. For four waves with branch indices $s_j=\pm1$, a resonant quartet must satisfy both frequency conservation and propagation-constant matching,}
\begin{align}
 \omega_1+\omega_2&=\omega_3+\omega_4,\notag\\
 \beta_{s_1}(\omega_1)+\beta_{s_2}(\omega_2)
 &=\beta_{s_3}(\omega_3)+\beta_{s_4}(\omega_4).
 \label{eq:beta-matching}
\end{align}
Let
\begin{equation*}
 \omega_{3,4}=\frac{\omega_1+\omega_2\pm\delta}{2},
 \qquad
 \Delta\omega=\omega_2-\omega_1.
\end{equation*}
Using Eq.~\eqref{eq:dispersion-branches}, the propagation-constant condition becomes
\begin{equation*}
 \delta^2=\Delta\omega^2-2\left[
 s_1\varepsilon(\omega_1)+s_2\varepsilon(\omega_2)
 -s_3\varepsilon(\omega_3)-s_4\varepsilon(\omega_4)
 \right].
\end{equation*}

\rev{The remainder of the paper focuses on degenerate FWM, because it gives the clearest setting in which branch structure and non-Hermitian mode composition can be varied independently of a pump-frequency difference. The two pump waves therefore have the same frequency and belong to the same branch,}
\begin{equation*}
 \omega_1=\omega_2\equiv\omega_p,
 \qquad s_1=s_2.
\end{equation*}
The generated sidebands are
\begin{equation*}
 \omega_{\pm}=\omega_p\pm\frac{\delta}{2},
\end{equation*}
and the matching equation reduces to
\begin{equation}
 \delta^2=-2\left[
 2s_1\varepsilon(\omega_p)
 -s_3\varepsilon\left(\omega_p+\frac{\delta}{2}\right)
 -s_4\varepsilon\left(\omega_p-\frac{\delta}{2}\right)
 \right].
 \label{eq:degenerate-phase-matching}
\end{equation}
\rev{Equation~\eqref{eq:degenerate-phase-matching} contains the entire linear resonance geometry of the degenerate process. When $s_3=s_4$, a valid separation occurs in the symmetric pair $\pm\delta$. When $s_3\neq s_4$, changing $\delta\to-\delta$ exchanges the generated frequencies and simultaneously maps $(s_3,s_4)$ into $(s_4,s_3)$. Thus the sign of the detuning by itself does not define a new physical channel; what matters is the combination of branch labels and generated frequencies.}

\rev{Although eight branch triples are possible at the level of notation, not all of them can satisfy Eq.~\eqref{eq:degenerate-phase-matching}. Within the unbroken-$\mathcal{CPT}$ and weak-dispersion domain specified above, four triples can be excluded analytically; Appendix~A gives the proof without relying on a numerical parameter scan. The four branch combinations that remain as possible resonant channels are}
\begin{table}[!htbp]
\centering
\caption{Allowed degenerate FWM branch configurations.}
\label{tab:configurations}
\begin{tabular}{cccc}
\toprule
Configuration & Pump $s_1$ & Sideband $s_3$ & Sideband $s_4$\\
\midrule
(I)   & $-$ & $+$ & $+$\\
(II)  & $-$ & $-$ & $+$\\
(III) & $-$ & $+$ & $-$\\
(IV)  & $+$ & $+$ & $+$\\
\bottomrule
\end{tabular}
\end{table}
\rev{Configurations (II) and (III) are related by exchanging the two generated sidebands and therefore have analogous matching structure. Configuration (IV) is qualitatively different: pump, signal, and idler all occupy the upper branch, so the resonance does not rely on an interbranch energy splitting. The numerical scans below show that this same-branch channel is not generic. It appears only in a narrow region close to the $\mathcal{CPT}$-breaking threshold, where the branch geometry changes rapidly and the eigenvectors are close to coalescence. This localization near the spectral degeneracy is one of the main dynamical distinctions between configuration (IV) and the three interbranch channels.}

\subsection{Polynomial sideband equation}

Set
\begin{equation*}
\begin{aligned}
 Q&=\frac{\delta^2}{4}, & K_p&=\widehat K(\omega_p),\\
 \varepsilon_p&=\varepsilon(\omega_p), & B&=2\kappa_2\omega_p+\kappa_1.
\end{aligned}
\end{equation*}
\rev{To locate the sideband separations efficiently over parameter space, we convert the square-root matching relation into an algebraic equation. Eliminating the square roots in Eq.~\eqref{eq:degenerate-phase-matching} gives the necessary condition}
\begin{equation}
 4Q\left(a_3Q^3+a_2Q^2+a_1Q+a_0\right)=0,
 \label{eq:cubic-Q}
\end{equation}
where
\begin{align*}
 a_0={}&B^2\gamma^2+2\varepsilon_p^2
 \left(s_1\varepsilon_p-\kappa_2K_p\right),\\
 a_1={}&-2B^2s_1\varepsilon_p+2B^2\kappa_2K_p
 -4s_1\kappa_2K_p\varepsilon_p
 +(5-\kappa_2^2)\varepsilon_p^2,\\
 a_2={}&B^2(\kappa_2^2-1)-2\kappa_2K_p
 +2s_1\varepsilon_p(2-\kappa_2^2),\\
 a_3={}&1-\kappa_2^2.
\end{align*}
\rev{The prefactor $Q=0$ represents the trivial zero-separation solution and is not a frequency-conversion channel. Positive real roots of the cubic provide candidate nonzero sideband separations, but the polynomial alone is deliberately not used as the final physical criterion. Repeated squaring can introduce extraneous roots, and the elimination procedure removes the explicit labels $s_3$ and $s_4$. Consequently every positive root is substituted back into the unsquared Eq.~\eqref{eq:degenerate-phase-matching}. This final test both rejects spurious algebraic solutions and assigns each surviving root to a definite branch configuration. The distinction is particularly important near the $\mathcal{CPT}$ threshold, where several candidate roots can lie close to one another.}

\begin{figure*}[t]
\centering
\includegraphics[width=0.48\textwidth]{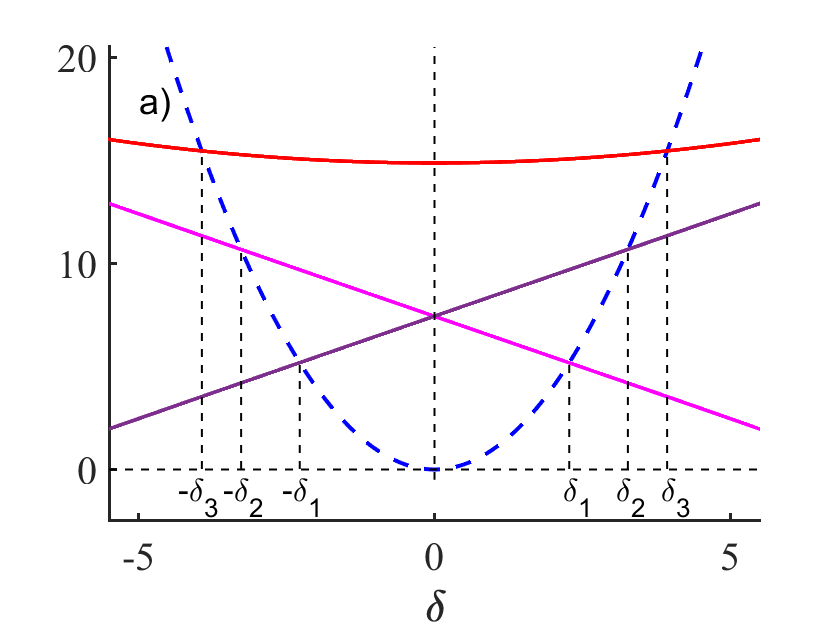}
\includegraphics[width=0.48\textwidth]{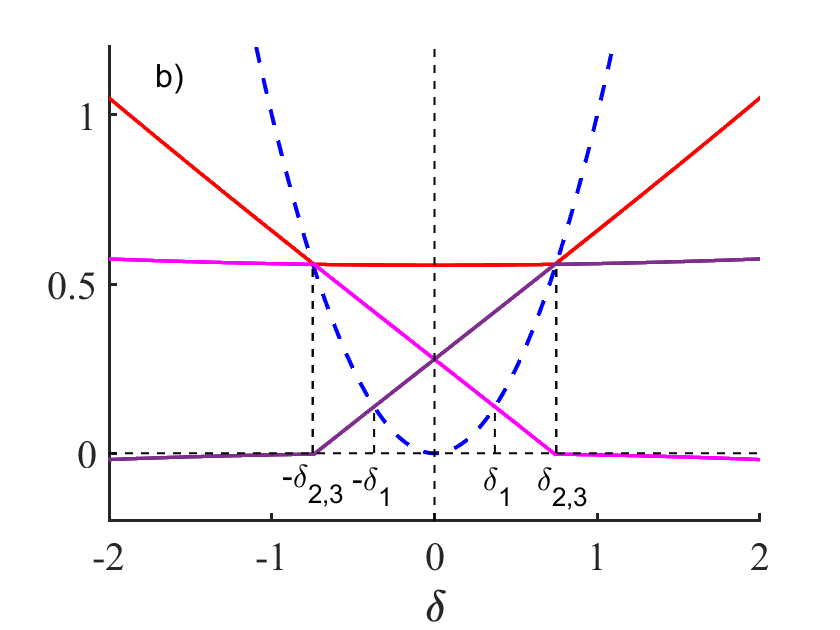}
\includegraphics[width=0.62\textwidth]{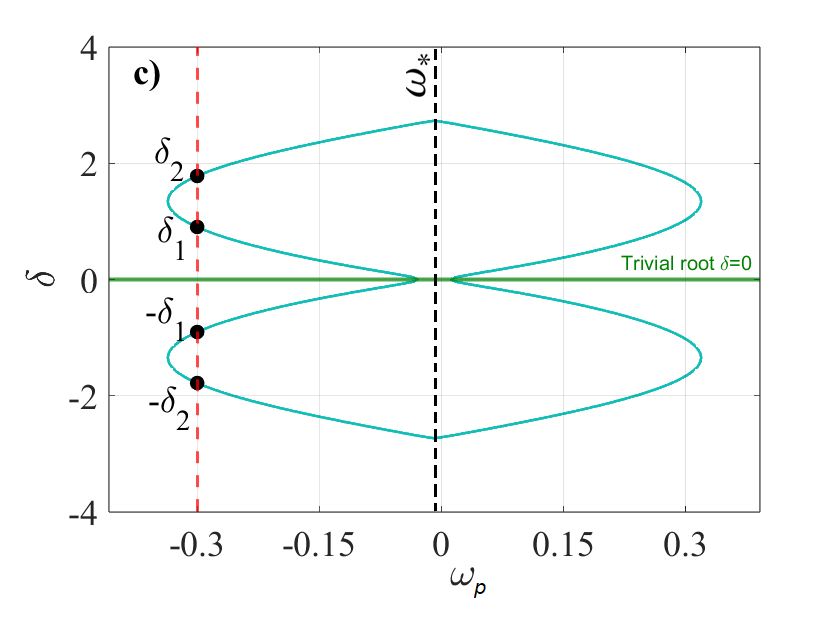}
\caption{Graphical solutions of Eq.~\eqref{eq:degenerate-phase-matching}. (a) Configurations (I)--(III) for $\{\kappa_0,\kappa_1,\kappa_2\}=\{1,0.3,0.04\}$, $\gamma=0.4$, and $\omega_p=2.3$. (b) The same three branch combinations at $\gamma=\gamma_{\mathcal{CPT}}=0.4375$ and $\omega_p=-3.38$. The blue dashed curves are the left-hand side and the solid curves are the configuration-dependent right-hand sides. (c) Zero contours for configuration (IV) with $\{\kappa_0,\kappa_1,\kappa_2\}=\{1,0.01,0.6\}$ and $\gamma=\gamma_{\mathcal{CPT}}-10^{-5}$. The green line is the trivial root $\delta=0$ and the cyan curves are nonzero solutions. At $\omega_p=-0.3$ (red dashed line), two positive roots and their negative counterparts coexist. The black dashed line marks $\omega_*=-\kappa_1/(2\kappa_2)$.}
\label{fig:phase-matching}
\end{figure*}

\rev{Figure~\ref{fig:phase-matching} illustrates how this resonance structure changes as the system approaches eigenmode coalescence. Panels (a) and (b) show the interbranch roots below and at the $\mathcal{CPT}$ threshold. The visible rearrangement of the intersections reflects the rapid change of both the branch separation and the internal eigenmode phase near the threshold. Panel (c) shows the more unusual intrabranch configuration (IV). At the marked pump frequency the original, unsquared matching equation possesses four nonzero roots, $\pm\delta_1$ and $\pm\delta_2$. The positive roots correspond to two distinct signal--idler separations for the same pump, so two degenerate FWM resonances coexist before nonlinear propagation is considered. This coexistence is the linear precursor of the simultaneous channels observed in the pulse simulations.}

\section{Group velocities and lack of Galilean invariance}

\rev{Phase matching determines whether coherent energy exchange is allowed, but finite pulses introduce an additional dynamical scale through their relative motion. In the frequency-independent coupler this issue is simplified by Galilean invariance: a common carrier-frequency shift can be removed by the usual boost of the nonlinear Schr\"odinger equation, which is why Ref.~\cite{wasak2015} could choose a zero pump frequency without loss of generality. The derivative terms in $\mathcal K$ destroy that equivalence. Under a transformation to a frame moving at velocity $v$,}
\begin{equation*}
 q_n(z,\tau)=e^{i(v\tau/2-v^2z/4)}w_n(z,\xi),
 \qquad \xi=\tau-vz,
\end{equation*}
the coupling operator becomes \cite{zezyulin2017}
\begin{equation*}
 \mathcal K_v=
 \kappa_0-\frac{v\kappa_1}{2}+\frac{v^2\kappa_2}{4}
 +i(\kappa_1-\kappa_2v)\partial_\xi
 -\kappa_2\partial_\xi^2.
\end{equation*}
\rev{The transformed operator shows explicitly that a boost changes the effective coupling coefficients themselves. Two phase-matched solutions with different carrier frequencies are therefore not equivalent descriptions of the same physical state in different frames. The pump frequency $\omega_p$ must be retained as an independent control parameter, as in spin--orbit-coupled systems \cite{hung2020}. This lack of Galilean invariance also explains why shifting the pump can alter not only the temporal trajectories but also the nonlinear conversion amplitude.}

\begin{figure*}[t]
\centering
\includegraphics[width=0.32\textwidth]{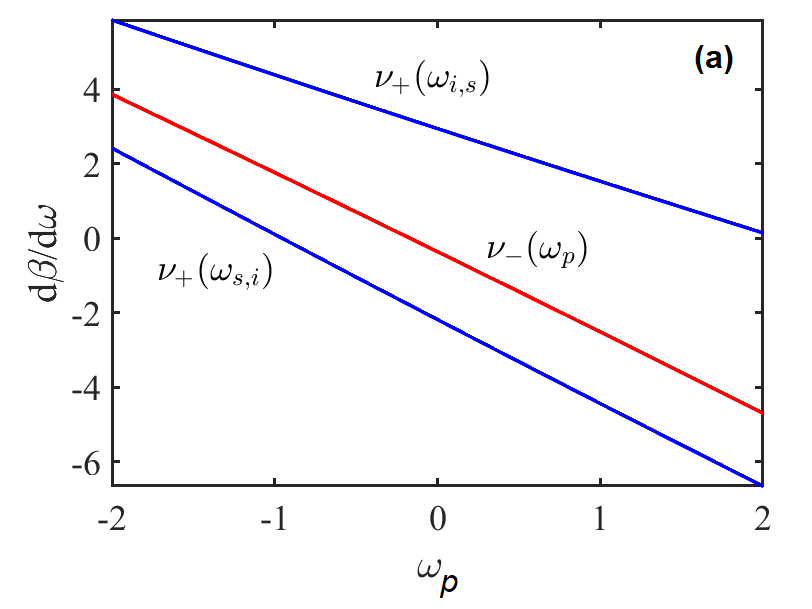}
\hfill
\includegraphics[width=0.32\textwidth]{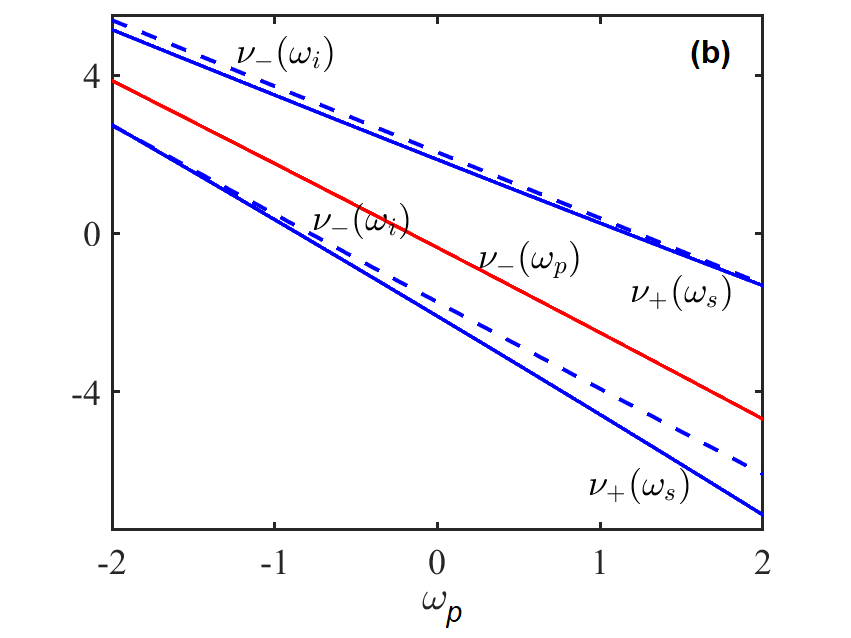}
\hfill
\includegraphics[width=0.32\textwidth]{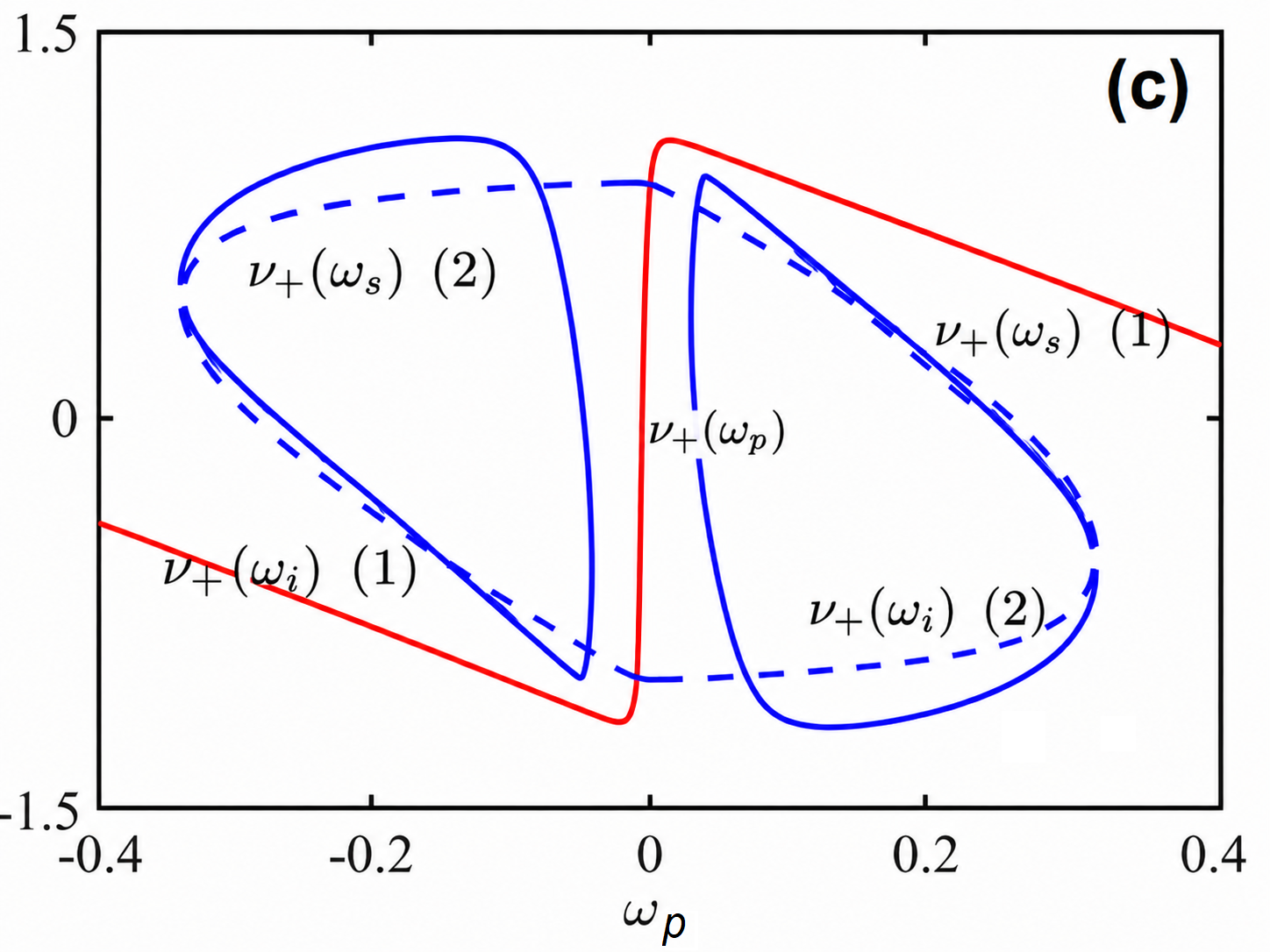}
\caption{Group velocities of the phase-matched waves as functions of $\omega_p$. (a) Configuration (I). (b) Configurations (II) and (III), shown by solid and dashed blue curves. Panels (a) and (b) use $\{\gamma,\kappa_1,\kappa_2\}=\{0.4,0.3,0.04\}$. (c) Configuration (IV) for $\{\gamma,\kappa_1,\kappa_2\}=\{\gamma_{\mathcal{CPT}}-10^{-5},0.01,0.6\}$. Red curves represent the pump branch and blue curves the generated sidebands. Crossings indicate exact group-velocity matching; nearby regions provide long but finite pulse overlap.}
\label{fig:group-velocities}
\end{figure*}

The group velocity in retarded-time coordinates is
\begin{equation}
 v_s(\omega)=\frac{d\beta_s}{d\omega}
 =-2\omega+s\frac{\widehat K(\omega)\left(2\kappa_2\omega+\kappa_1\right)}
 {\sqrt{\widehat K^2(\omega)-\gamma^2}}.
 \label{eq:group-velocity}
\end{equation}
\rev{Group velocity is not an additional FWM conservation law, and exact velocity matching is not required for the plane-wave resonance. It nevertheless controls how long finite packets remain overlapped and can therefore exchange energy coherently. A large mismatch shortens the interaction length and suppresses conversion even when the phase-matching condition is exact. Conversely, nearly equal velocities increase the overlap length, although they can make the generated packets difficult to separate in the time domain. For this reason the numerical examples are selected using both the kinematic phase-matching equation and the dynamical overlap information contained in Eq.~\eqref{eq:group-velocity}. The crossings and near crossings in Fig.~\ref{fig:group-velocities} identify pump frequencies where these two requirements can be satisfied simultaneously over an appreciable propagation distance.}

\section{Results and discussion}

\subsection{Pulse simulations and spectral convention}

\rev{The analytical matching calculation treats monochromatic modes, whereas the actual conversion process is tested with finite pulses. We propagate the coupled equations with a Fourier pseudospectral scheme, evaluating the linear two-core step in frequency space and the local Kerr response in retarded time. This separation is convenient because the dispersive coupling is diagonal in frequency while the Kerr term is local in retarded time. The input is a superposition of Gaussian wave packets,}
\begin{equation}
 \mathbf q(0,\tau)=e^{-\tau^2/(4T_0^2)}
 \sum_{j=1}^{3}A_j\mathbf r_{s_j}(\omega_j)e^{-i\omega_j\tau},
 \label{eq:gaussian-input}
\end{equation}
\rev{where one term is the pump, one of the phase-matched sidebands is weakly seeded, and the conjugate sideband is initially absent. Unless stated otherwise, $T_0=40$. This choice provides a narrow spectrum while retaining a finite temporal extent, so walk-off can be observed directly during propagation. The carrier frequencies are always taken from the original unsquared matching equation rather than from the polynomial alone. The simulations therefore test the same branch-resolved resonances established analytically, while allowing spectral broadening, finite pulse overlap, and additional nonlinear products to emerge dynamically.}

The numerical spectra use
\begin{equation*}
 \widetilde q(\Omega)=\int_{-\infty}^{\infty}q(\tau)e^{-i\Omega\tau}d\tau.
\end{equation*}
\rev{Because the analytical modal ansatz contains $e^{-i\omega\tau}$ while the numerical Fourier transform above uses the kernel $e^{-i\Omega\tau}$, a theoretical mode labeled by $\omega$ appears in the plotted spectra at $\Omega=-\omega$. This sign reversal is purely a plotting convention and does not change the matching relations. We retain the convention used in the numerical data and explicitly distinguish the displayed coordinate $\Omega$ from the analytical frequency $\omega$ throughout the discussion.}

\subsection{Interbranch conversion}

\begin{figure*}[t]
\centering
\includegraphics[width=0.96\textwidth]{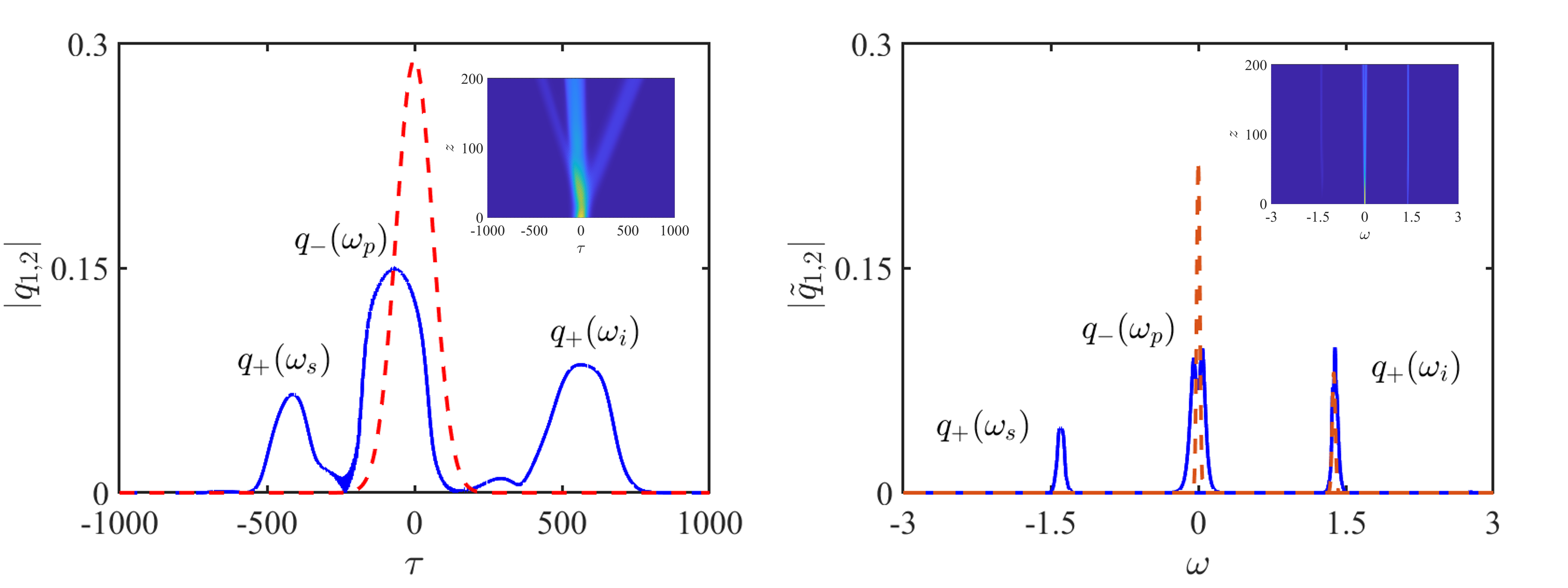}
\caption{Configuration (I) for $\{\kappa_0,\kappa_1,\kappa_2\}=\{1,0.3,0.04\}$, $\gamma=0.4$, $\sigma=+1$, and $\omega_p=0$. The pump and seed amplitudes are $A_p=0.24$ and $A_s=0.11$. Left: initial (red dashed) and output (blue) temporal amplitude profiles; the inset shows propagation. Right: corresponding spectra in the numerical coordinate $\Omega=-\omega$. The phase-matched idler grows while the packets separate according to their group velocities.}
\label{fig:config-I-zero-pump}
\end{figure*}

\rev{Configuration (I) provides the simplest benchmark because both generated waves lie on the upper branch while the pump occupies the lower branch. At $\omega_p=0$, one upper-branch sideband is weakly seeded [Fig.~\ref{fig:config-I-zero-pump}], and the conjugate upper-branch idler grows at the frequency predicted by Eq.~\eqref{eq:degenerate-phase-matching}. The temporal trajectories of the seeded and generated packets follow the branch-dependent group velocities from Eq.~\eqref{eq:group-velocity}. The packets therefore separate gradually during propagation, but the overlap remains long enough for a clearly resolved idler to develop. This agreement between the predicted resonance, the packet trajectories, and the output spectrum establishes a reference case before the pump is shifted away from zero.}

\begin{figure*}[t]
\centering
\includegraphics[width=0.96\textwidth]{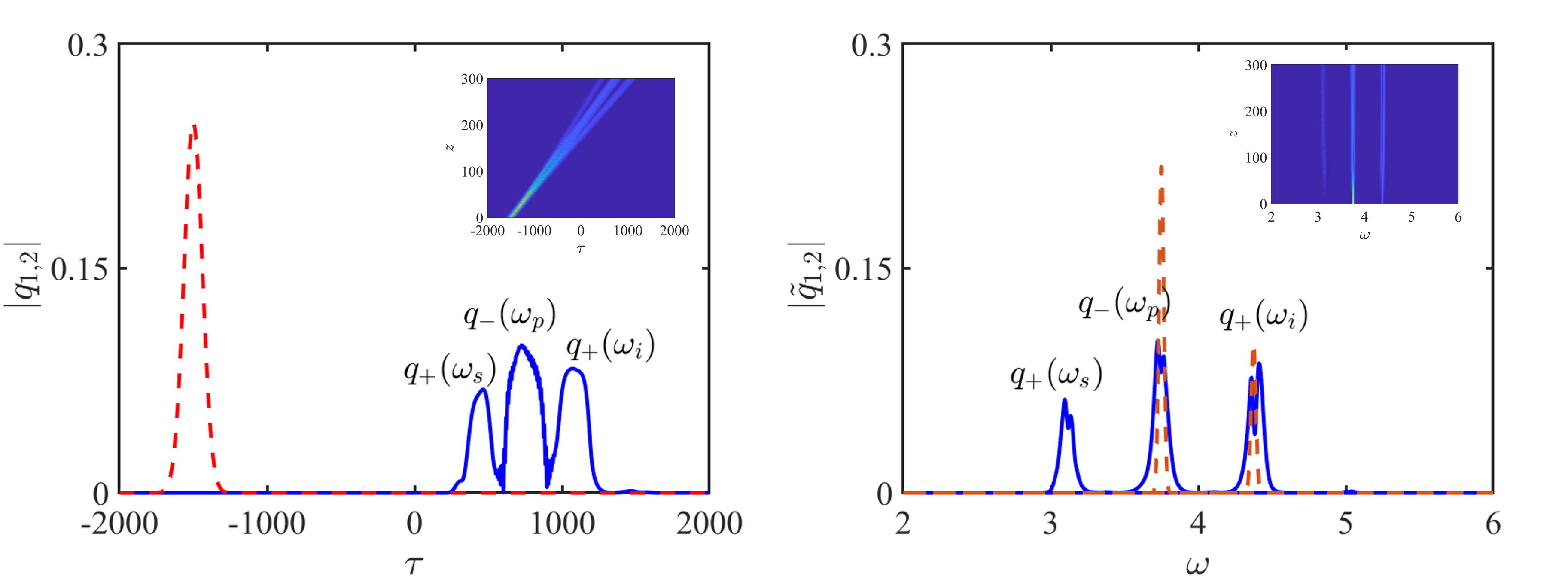}
\caption{Configuration (I) with the parameters of Fig.~\ref{fig:config-I-zero-pump} except $\omega_p=-3.75$. The substantially different group velocities produce a much larger temporal displacement. The spectral coordinate is $\Omega=-\omega$, so the pump appears near $\Omega=3.75$.}
\label{fig:config-I-shifted-pump}
\end{figure*}

\rev{The calculation at $\omega_p=-3.75$ [Fig.~\ref{fig:config-I-shifted-pump}] demonstrates why the loss of Galilean invariance is dynamically relevant rather than merely formal. It is not a frequency-translated copy of the zero-pump case: the phase-matched sidebands shift, the eigenvector phases change, the nonlinear overlaps are modified, and the three packets acquire different group velocities. The larger walk-off is visible as a pronounced output delay and shortens the distance over which the waves interact strongly. The resulting conversion amplitude therefore changes together with the temporal geometry, even though the process is still described by the same branch configuration.}

\begin{figure*}[t]
\centering
\includegraphics[width=0.96\textwidth]{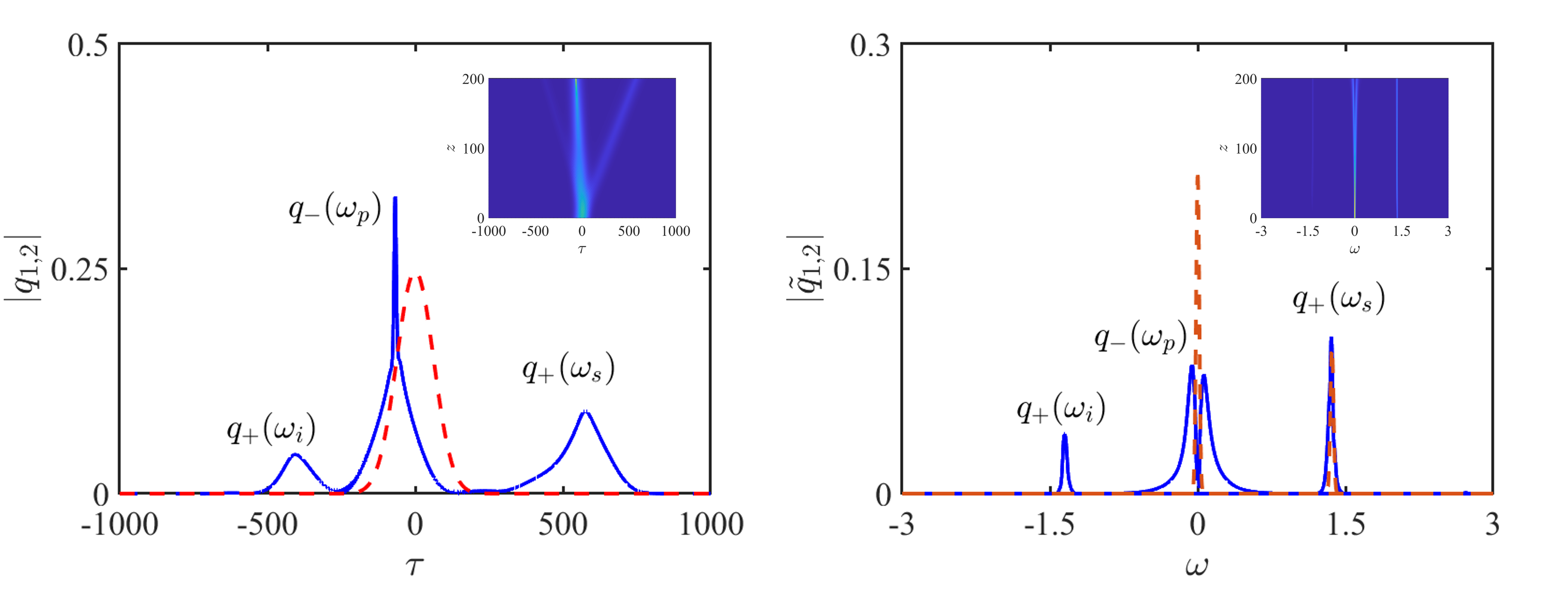}
\caption{Configuration (I) at $\gamma=\gamma_{\mathcal{CPT}}=0.4375$ with the remaining parameters as in Fig.~\ref{fig:config-I-zero-pump} and self-focusing nonlinearity $\sigma=-1$. Phase-matched sidebands are still generated, but the dynamics is more sensitive to modal nonorthogonality near the threshold.}
\label{fig:config-I-threshold}
\end{figure*}

\rev{Figure~\ref{fig:config-I-threshold} probes the same interbranch resonance at the $\mathcal{CPT}$ threshold and with the opposite sign of the Kerr coefficient. The linear roots remain phase matched because the kinematic condition is independent of the sign of $\sigma$. What changes is the nonlinear phase accumulation and therefore the detailed rate and phase of the energy exchange along $z$. The example also shows that phase matching alone does not determine the conversion history: close to the spectral threshold, modal nonorthogonality and nonlinear phase evolution become increasingly important even when the resonant frequencies themselves remain well defined.}

\begin{figure*}[t]
\centering
\includegraphics[width=0.96\textwidth]{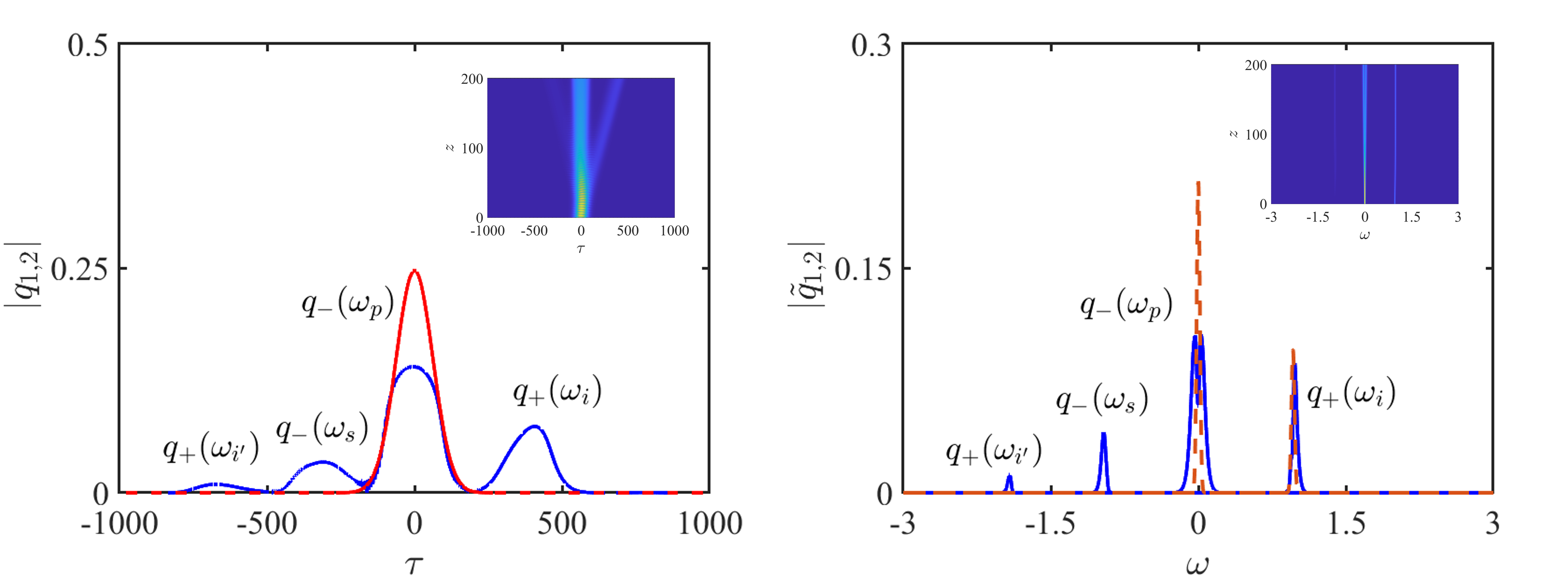}
\caption{Configuration (II), $2\mathbf q_-\rightarrow\mathbf q_-+\mathbf q_+$, for $\{\kappa_0,\kappa_1,\kappa_2\}=\{1,0.01,0.001\}$, $\gamma=0.4$, $\sigma=+1$, and $\omega_p=0$. The input amplitudes are $A_p=0.24$ and $A_s=0.11$. A weak fifth spectral component, labeled $\omega_i'$, is generated by a secondary resonant interaction. The displayed spectral coordinate is $\Omega=-\omega$.}
\label{fig:config-II}
\end{figure*}

\rev{Configuration (II) mixes one lower- and one upper-branch sideband, while configuration (III) is obtained by exchanging those generated waves. Figure~\ref{fig:config-II} shows that the primary seeded-idler pair is not the end of the nonlinear evolution. Once the initially phase-matched waves have acquired appreciable amplitude, a weaker fifth spectral component appears. The same qualitative mechanism was identified in the nondispersive $\mathcal{PT}$ coupler \cite{wasak2015}: the first resonant quartet creates new populated frequencies, and these can in turn participate in a second quartet that also satisfies the frequency and branch constraints. In the present dispersive system, the secondary resonance is additionally influenced by the frequency dependence of the eigenvectors and coupling, so it provides an early indication that a strict three-wave truncation cannot describe the long-distance dynamics.}

\subsection{Intrabranch mixing near the \texorpdfstring{$\mathcal{CPT}$}{CPT} threshold}

\begin{figure*}[t]
\centering
\includegraphics[width=0.94\textwidth]{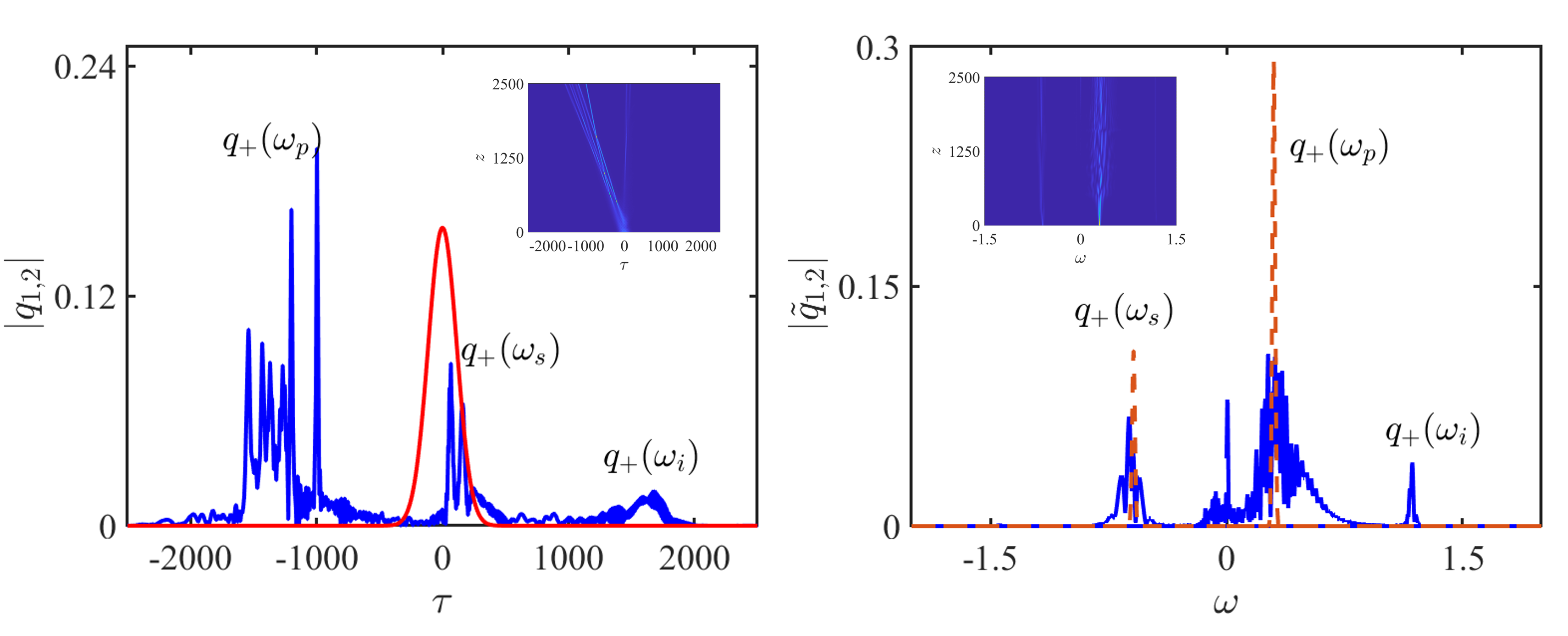}
\caption{Configuration (IV), $2\mathbf q_+\rightarrow\mathbf q_++\mathbf q_+$, for $\{\kappa_0,\kappa_1,\kappa_2\}=\{1,0.01,0.6\}$, $\gamma=\gamma_{\mathcal{CPT}}-10^{-5}$, $\omega_p=-0.3$, and $\sigma=+1$. The input amplitudes are $A_p=0.16$ and $A_s=0.06$. The main phase-matched peaks are accompanied by a broad set of secondary components because the branch eigenvectors are nearly coalescent. The displayed spectral coordinate is $\Omega=-\omega$.}
\label{fig:config-IV}
\end{figure*}

\rev{Configuration (IV) is dynamically the most distinctive case because it is confined to the neighborhood of eigenmode coalescence. In Fig.~\ref{fig:config-IV}, the strongest spectral peaks still occur at the frequencies selected by the unsquared phase-matching equation, confirming that the linear resonance remains a useful organizing principle. The surrounding spectrum, however, is much richer than in configurations (I)--(III): many secondary components appear during propagation. In this parameter range $\cos\phi(\omega)$ is small over part of the relevant spectrum, so the biorthogonal normalization becomes poorly conditioned and the distinction between the two linear eigenmodes is fragile. The phase-matching equation therefore identifies the dominant resonances but should not be interpreted as a statement that the nonlinear evolution is restricted to only three populated modes.}

\begin{figure*}[t]
\centering
\includegraphics[width=0.96\textwidth]{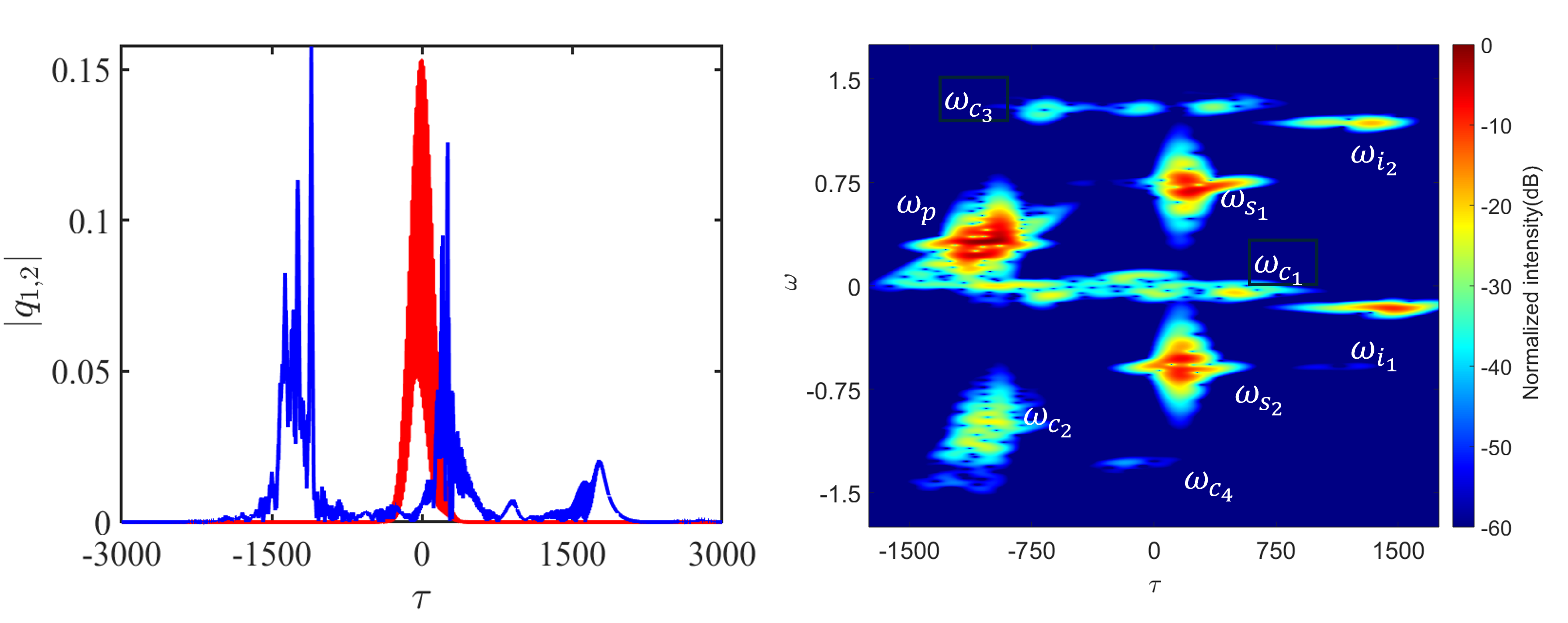}
\caption{Two simultaneous configuration-(IV) channels with the parameters of Fig.~\ref{fig:config-IV}. The pump amplitude is $A_p=0.12$ and the two seed amplitudes are $A_{s1}=A_{s2}=0.05$. Left: initial state at $z=0$ (red dashed) and output at $z=3000$ (blue). Right: output spectrogram. The spectral axis is the numerical coordinate $\Omega=-\omega$; the principal peaks are identified in Table~\ref{tab:spectral-peaks}.}
\label{fig:two-FWM}
\end{figure*}

\rev{The near-threshold resonance structure also permits coexistence rather than merely stronger single-channel conversion. For the same parameters, Eq.~\eqref{eq:degenerate-phase-matching} has two positive sideband separations, $\delta_1\simeq0.9033$ and $\delta_2\simeq1.7809$. They correspond to two distinct signal--idler pairs associated with the same pump and the same upper branch. Seeding one sideband from each pair therefore initiates two FWM channels simultaneously, as shown in Fig.~\ref{fig:two-FWM}. This is the time-domain realization of the multiple roots already visible in Fig.~\ref{fig:phase-matching}(c).}

The short-time Fourier transform is
\begin{equation*}
 S(\tau,\Omega)=\left|
 \int_{-\infty}^{\infty}q(z,t)h(t-\tau)e^{-i\Omega t}dt
 \right|^2,
\end{equation*}
\rev{where $h$ is a localized analysis window. The short-time spectrum is used here only as a diagnostic of how different frequency components separate in retarded time; it does not enter the propagation algorithm. Table~\ref{tab:spectral-peaks} separates the four targeted sidebands from the additional products created by cascading. The central feature near $\Omega=0$ is comparatively broad in both time and frequency and contains overlapping contributions, so it cannot be assigned reliably to a single isolated linear mode from the spectrogram alone. The coexistence of the targeted peaks with these additional components illustrates the transition from a few-resonance picture to a genuinely multifrequency nonlinear dynamics near the exceptional point.}

\begin{table}[!htbp]
\centering
\caption{Spectral peaks in Fig.~\ref{fig:two-FWM}. The displayed coordinate is $\Omega=-\omega$.}
\label{tab:spectral-peaks}
\renewcommand{\arraystretch}{1.15}
\begin{tabular}{ccc}
\toprule
$\Omega$ & Assignment & Temporal interval $\tau$\\
\midrule
$\sim 1.272$  & secondary component & $[-1100,1100]$\\
$1.1905$       & idler 2              & $[-930,1854]$\\
$0.7517$       & signal 1             & $[-95,780]$\\
$0.3000$       & pump                 & $[-2170,-820]$\\
$\sim0.00733$  & overlapping components & $[-2106,1449]$\\
$-0.1517$      & idler 1              & $[1135,2202]$\\
$-0.5905$      & signal 2             & $[-201.2,570]$\\
$\sim-1.1905$ & secondary component  & $[-1610,-975]$\\
$\sim-1.288$  & secondary component  & $[-650,225.4]$\\
\bottomrule
\end{tabular}
\end{table}

\subsection{Plane-wave approximation}

\rev{To distinguish the basic resonant exchange from the effects of finite bandwidth and pulse separation, we now introduce a reduced three-wave model. Its purpose is not to replace the full simulations over arbitrarily long distances, but to identify the leading nonlinear coupling among the pump, seed, and generated idler while those components remain narrow in frequency and overlapped in time. The construction follows Ref.~\cite{wasak2015}, with an essential modification: projection is performed with the frequency-dependent dual modes of the dispersive non-Hermitian coupler. Finite bandwidth, walk-off, spectral broadening, and secondary resonances are intentionally omitted. Their influence can therefore be diagnosed by comparing the reduced dynamics with the full pulse propagation.}

Let $\omega_1=\omega_p$ be the pump frequency and let $\omega_3$ and $\omega_4$ denote the seeded and generated sidebands, respectively. Under exact frequency and propagation-constant matching, we write the field as a superposition of three linear modes,
\begin{align}
 \mathbf q(z,\tau)={}&2a_1(z)e^{is_1\phi_1}
 \mathbf r_{s_1}(\omega_1)e^{i\beta_1z-i\omega_1\tau}
 \nonumber\\
 &+a_3(z)\mathbf r_{s_3}(\omega_3)e^{i\beta_3z-i\omega_3\tau}
 \nonumber\\
 &+a_4(z)\mathbf r_{s_4}(\omega_4)e^{i\beta_4z-i\omega_4\tau},
 \label{eq:three-wave-ansatz}
\end{align}
\rev{where $\phi_j=\phi(\omega_j)$, $\beta_j=\beta_{s_j}(\omega_j)$, and the amplitudes $a_j(z)$ vary slowly compared with the carrier phases. The factor 2 in the pump term is a normalization choice convenient for degenerate FWM and leads to $u_1=4|a_1|^2$ below. The branch-dependent phase factor $e^{is_1\phi_1}$ is likewise introduced only to simplify the final resonant coefficients; it does not modify the physical field represented by the ansatz. Writing the expansion in the linear eigenmode basis makes the connection between the phase-matching analysis and the nonlinear energy exchange explicit.}

\rev{Substitution of Eq.~\eqref{eq:three-wave-ansatz} into the Kerr terms separates the nonlinear response into self- and cross-phase-modulation contributions and three products that carry the phases required for resonant FWM: $a_1^*a_3a_4$, $a_1^2a_4^*$, and $a_1^2a_3^*$. Terms whose longitudinal or temporal phases do not match are rapidly oscillating and are dropped in the rotating-wave approximation. The linear contribution cancels exactly because each carrier factor $\mathbf r_{s_j}(\omega_j)e^{i\beta_jz-i\omega_j\tau}$ already satisfies the linear eigenvalue problem. The approximation therefore isolates the resonant nonlinear exchange without altering the underlying non-Hermitian mode structure.}

\rev{At each carrier frequency the resonant nonlinear source is a two-component vector in the core basis. To extract the amplitude driving a chosen branch, we project this vector with $\boldsymbol\ell_{s_j}^{\dagger}$ from Eq.~\eqref{eq:left-eigenvector}. Biorthogonality removes the opposite branch and leaves one scalar evolution equation for each $a_j(z)$. This step is where the non-Hermitian character enters the reduced nonlinear model most directly: using a conventional Hermitian projection would give incorrect overlap factors when the right eigenvectors are nonorthogonal. The resulting amplitude system is}
\begin{equation}
\begin{aligned}
 i\frac{da_1}{dz}&=\sigma\left(S_1a_1+\Lambda_1a_1^*a_3a_4\right),\\
 i\frac{da_3}{dz}&=\sigma\left(S_3a_3+\Lambda_3a_1^2a_4^*\right),\\
 i\frac{da_4}{dz}&=\sigma\left(S_4a_4+\Lambda_4a_1^2a_3^*\right).
\end{aligned}
\label{eq:amplitude-system}
\end{equation}
where
\begin{equation*}
\begin{aligned}
 S_1&=\frac{u_1}{2}+u_3+u_4,\\
 S_3&=u_1+\frac{u_3}{2}+u_4,\\
 S_4&=u_1+u_3+\frac{u_4}{2}.
\end{aligned}
\end{equation*}
and
\begin{equation*}
 u_1=4|a_1|^2,\qquad u_3=|a_3|^2,\qquad u_4=|a_4|^2.
\end{equation*}
\rev{The real quantities $S_j$ describe the self- and cross-phase shifts produced by the retained waves. The coefficients $\Lambda_j$, by contrast, govern the resonant transfer among the three carriers. Because they depend on the branch combination and on the internal phases $\phi(\omega_j)$, they need not be real and they need not be equal for the three amplitude equations. This frequency dependence is the central difference from the nondispersive $\mathcal{PT}$ coupler of Ref.~\cite{wasak2015}: coupling dispersion makes the nonlinear interaction itself sensitive to where the three frequencies lie on the two dispersion branches.}

\paragraph{Configuration (I): $2\mathbf q_-(\omega_1)\rightarrow \mathbf q_+(\omega_3)+\mathbf q_+(\omega_4)$.}
For this process, direct substitution and projection give
\begin{align}
 \Lambda_1^{\mathrm I}&=\frac{e^{i\phi_1}\left[1+e^{-i(\phi_3+\phi_4)}\right]}{2\cos\phi_1},\nonumber\\
 \Lambda_3^{\mathrm I}&=\frac{e^{i\phi_4}+e^{i(\phi_3-2\phi_1)}}{\cos\phi_3},\nonumber\\
 \Lambda_4^{\mathrm I}&=\frac{e^{i\phi_3}+e^{i(\phi_4-2\phi_1)}}{\cos\phi_4}.
 \label{eq:lambda-I}
\end{align}
\rev{For frequency-independent coupling, $\phi_1=\phi_3=\phi_4$, so $\Lambda_1^{\mathrm I}=1$ and $\Lambda_3^{\mathrm I}=\Lambda_4^{\mathrm I}=2$. In this limit the phase dependence drops out of the resonant coefficients, Eq.~\eqref{eq:amplitude-system} reduces to the conservative FWM model of Ref.~\cite{wasak2015}, and the selected modal power satisfies}
\begin{equation*}
 \frac{d}{dz}(u_1+u_3+u_4)=0.
\end{equation*}
\rev{With dispersive coupling, the three phases are generally different. The coefficients $\Lambda_j^{\mathrm I}$ then become unequal complex numbers, so the reduced resonant subsystem no longer possesses this simple modal-power conservation law. This change is not caused by additional waves in the truncation; it arises already at the level of three perfectly phase-matched modes because the non-Hermitian projection varies with frequency.}

\paragraph{Configurations (II) and (III).}
For configuration (II), $2\mathbf q_-(\omega_1)\rightarrow \mathbf q_-(\omega_3)+\mathbf q_+(\omega_4)$, the coefficients are
\begin{align}
 \Lambda_1^{\mathrm{II}}&=\frac{e^{i\phi_1}\left[1-e^{i(\phi_3-\phi_4)}\right]}{2\cos\phi_1},\nonumber\\
 \Lambda_3^{\mathrm{II}}&=\frac{e^{-i(2\phi_1+\phi_3)}-e^{i\phi_4}}{\cos\phi_3},\nonumber\\
 \Lambda_4^{\mathrm{II}}&=\frac{e^{i(\phi_4-2\phi_1)}-e^{-i\phi_3}}{\cos\phi_4}.
 \label{eq:lambda-II}
\end{align}
Configuration (III) follows by interchanging the generated waves, $3\leftrightarrow4$. In the nondispersive limit, $\Lambda_1^{\mathrm{II}}=\Lambda_4^{\mathrm{II}}=0$, whereas
\begin{equation*}
 \Lambda_3^{\mathrm{II}}=-4i\sin\phi\,e^{-i\phi}.
\end{equation*}
\rev{In the nondispersive limit the vanishing of $\Lambda_1^{\mathrm{II}}$ and $\Lambda_4^{\mathrm{II}}$ means that the pump and the fast seeded wave acquire only phase modulation within the reduced model, while the generated slow wave can change its power through the gain--loss-assisted resonant term. This is the $\mathcal{PT}$-induced channel identified in Ref.~\cite{wasak2015}. Coupling dispersion removes that special algebraic simplification: $\phi_1$, $\phi_3$, and $\phi_4$ are different, all three resonant coefficients can become nonzero, and the generated-wave dynamics is coupled back to the other retained amplitudes. The resulting power exchange is therefore more symmetric dynamically even though the underlying gain--loss system remains nonconservative.}

\paragraph{Configuration (IV): $2\mathbf q_+(\omega_1)\rightarrow \mathbf q_+(\omega_3)+\mathbf q_+(\omega_4)$.}
For the intrabranch process, one obtains
\begin{align}
 \Lambda_1^{\mathrm{IV}}&=\frac{e^{-i\phi_1}\left[1+e^{-i(\phi_3+\phi_4)}\right]}{2\cos\phi_1},\nonumber\\
 \Lambda_3^{\mathrm{IV}}&=\frac{e^{i\phi_4}+e^{i(2\phi_1+\phi_3)}}{\cos\phi_3},\nonumber\\
 \Lambda_4^{\mathrm{IV}}&=\frac{e^{i\phi_3}+e^{i(2\phi_1+\phi_4)}}{\cos\phi_4}.
 \label{eq:lambda-IV}
\end{align}
\rev{In the parameter range explored here, configuration (IV) occurs only in a narrow region close to the $\mathcal{CPT}$-breaking threshold, and the reduced coefficients make explicit why this regime is delicate. As the exceptional point is approached, $\cos\phi_j\to0$, the $c$-norm $\mathbf r_{s_j}^T\mathbf r_{s_j}$ vanishes, and the two eigenvectors coalesce. The dual-mode normalization in Eq.~\eqref{eq:left-eigenvector} therefore becomes singular and the modal projection is ill-conditioned \cite{ruter2010,konotop2016}. The large values of the reduced coefficients should not be interpreted as evidence that the three-wave model becomes more accurate or that conversion simply becomes stronger. They are a symptom of the same near-degeneracy that makes the eigenbasis poorly conditioned and allows additional resonant and nonresonant components to participate. The rapid departure from the reduced dynamics is therefore a structural consequence of approaching the exceptional point, not merely a finite-pulse correction.}

For all configurations, Eq.~\eqref{eq:amplitude-system} implies the power balances
\begin{align}
 \frac{du_1}{dz}&=8\sigma\operatorname{Im}
 \left(\Lambda_1a_1^{*2}a_3a_4\right),\nonumber\\
 \frac{du_3}{dz}&=2\sigma\operatorname{Im}
 \left(\Lambda_3a_1^2a_3^*a_4^*\right),\nonumber\\
 \frac{du_4}{dz}&=2\sigma\operatorname{Im}
 \left(\Lambda_4a_1^2a_3^*a_4^*\right).
 \label{eq:reduced-power-balance}
\end{align}
\rev{The three right-hand sides do not generally cancel because the overlap coefficients are distinct complex numbers. Even within the reduced subsystem, the retained waves can therefore exchange power with the gain--loss background as well as redistribute power among one another. This distinguishes the present dynamics from a conservative three-wave interaction, where the resonant terms can be organized into simple power-balance relations. Conservative exchange is recovered only in special limits, including configuration (I) with frequency-independent coupling. The power-balance equations thus provide a compact way to see how non-Hermiticity enters the nonlinear dynamics after the linear resonance condition has already been imposed.}

\begin{figure*}[t]
\centering
\includegraphics[width=0.96\textwidth]{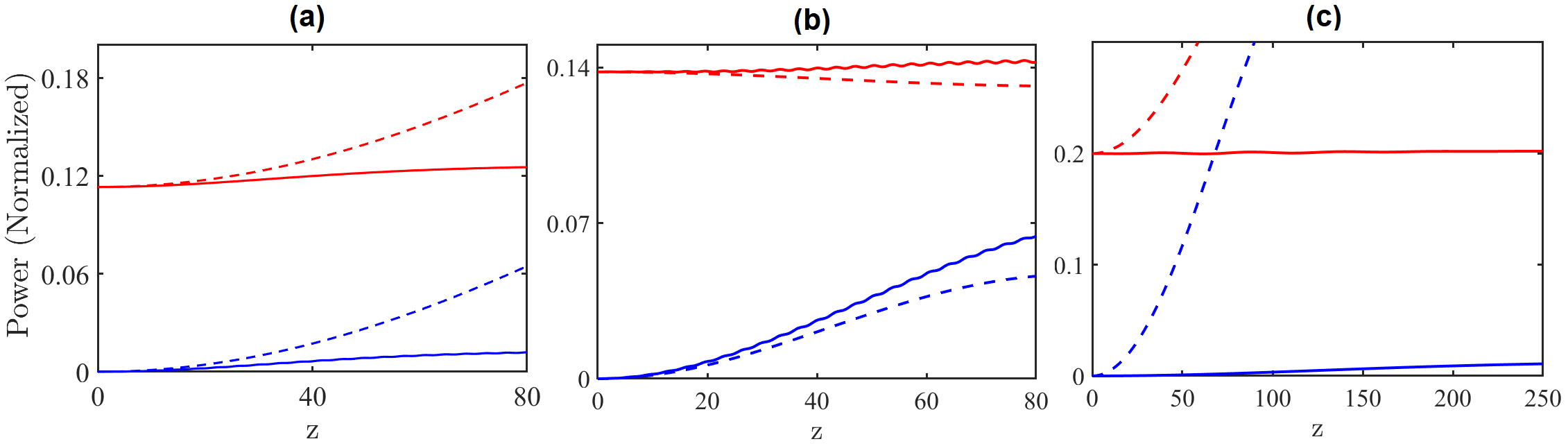}
\caption{Comparison of the full pulse simulations (solid curves) and the plane-wave model (dashed curves): (a) configuration (I), (b) configuration (II), and (c) configuration (IV). Red and blue denote the normalized seeded-sideband and generated-idler powers, respectively. The plane-wave approximation reproduces the initial resonant exchange for configurations (I) and (II). Deviations at longer distances result from finite bandwidth, temporal walk-off, nonlinear broadening, and secondary resonances. Near the exceptional point in configuration (IV), additional modes grow from the outset and the three-wave approximation is no longer sufficient.}
\label{fig:reduced-comparison}
\end{figure*}

\begin{figure*}[t]
\centering
\includegraphics[width=0.49\linewidth]{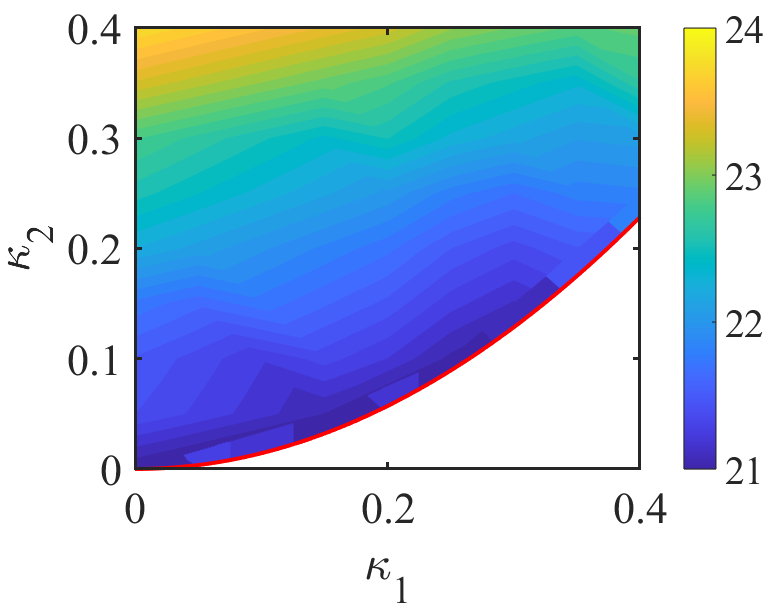}
\hfill
\includegraphics[width=0.47\linewidth]{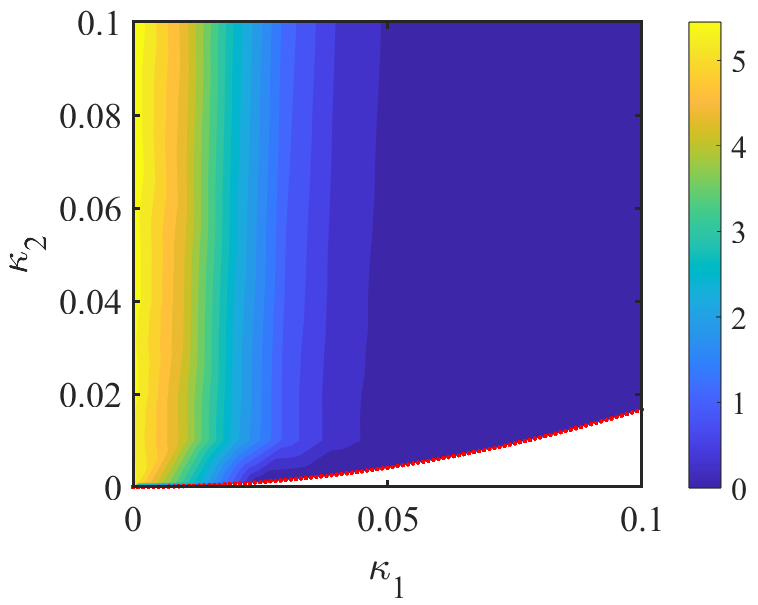}
\caption{Output FWM fraction $\eta_{\mathrm{FWM}}$ in the $(\kappa_1,\kappa_2)$ plane for self-defocusing nonlinearity, $\sigma=+1$. (a) Configuration (I), with $\omega_p=0$ and $\gamma=0.3$. (b) Configuration (II), with $\omega_p=0$ and $\gamma=0.4$. The color scales give the percentage of output power contained in the generated FWM sideband. The solid red curve is the exact all-real-spectrum boundary. 
The uncolored region below the red curve lies in the broken-spectrum domain.}
\label{fig:efficiency-maps}
\end{figure*}
\rev{For configurations (I) and (II), the dashed curves in Fig.~\ref{fig:reduced-comparison} follow the initial transfer while the packets are still overlapped and spectrally narrow. The later separation between the full and reduced calculations has a clear physical origin: temporal walk-off shortens the effective interaction, nonlinear broadening spreads power away from the carrier frequencies, and secondary resonances populate modes that are absent from the three-wave ansatz. Configuration (IV) departs much earlier. This early failure is consistent with the dense sideband structure in Figs.~\ref{fig:config-IV} and \ref{fig:two-FWM} and with the singular behavior of the dual basis near eigenmode coalescence. In this sense the reduced model is useful not only when it succeeds; the location of its breakdown also identifies a regime in which the full nonlinear dynamics can no longer be organized around a well-conditioned three-mode basis.}

\subsection{Conversion-efficiency maps}

\rev{To summarize the conversion over a wider parameter range, we use a quantity that can be extracted directly from the output spectrum of the full simulations. At the output distance $z_f$, the spectral fraction carried by the generated FWM sideband is defined as}
\begin{equation}
 \eta_{\mathrm{FWM}}(z_f)=
 \frac{P_{\mathrm{FWM}}(z_f)}{P_{\mathrm{tot}}(z_f)}\times100\%,
 \label{eq:FWM-efficiency}
\end{equation}
\rev{where $z_f$ is the output propagation distance. The generated-wave power $P_{\mathrm{FWM}}$ is obtained by integrating the spectral intensity over a narrow window centered on the generated idler peak, while $P_{\mathrm{tot}}$ is evaluated from the complete output spectrum at the same distance. Because Eq.~\eqref{eq:total-power-balance} allows the field to exchange power with the gain--loss background, Eq.~\eqref{eq:FWM-efficiency} should be interpreted as an output spectral fraction rather than as a conserved photon-conversion efficiency. The distinction is important when comparing different points in parameter space: a larger fraction can reflect both redistribution among frequencies and a different net gain--loss history. The integration window is therefore chosen only after the generated peak is spectrally resolved from the pump and the seeded sideband.}

\rev{The two sampled maps display noticeably different dynamical sensitivities. Configuration (I) retains an appreciable generated fraction over a relatively broad region, with the fraction generally increasing as $\kappa_2$ grows and decreasing as $\kappa_1$ becomes larger. Configuration (II) is more selective: its strongest response is concentrated at small $\kappa_1$, while the dependence on $\kappa_2$ is weaker away from the lower edge of the scan. This contrast is consistent with two mechanisms already identified above. The branch-dependent overlap coefficients respond differently to changes of the eigenvector phases, and the group-velocity mismatch changes the finite interaction length of the pulses. The maps should therefore be read as a combined dynamical response rather than as a phase-matching diagram alone. The hatched strip in each panel belongs to the unbroken-$\mathcal{CPT}$ domain but was not included in the original numerical grid, so no efficiency values are inferred there.}

\section{Conclusions}

\rev{The results can be summarized as a hierarchy linking the linear spectrum, resonant kinematics, and nonlinear dynamics. Coupling dispersion first reshapes the two-branch spectrum by fixing the minimum of $\widehat K(\omega)$ and therefore the boundary of the all-real $\mathcal{CPT}$ phase. The same frequency dependence removes the Galilean equivalence between different pump frequencies and makes the modal composition vary across a resonant quartet. Within the normalized weak-dispersion domain considered here, Appendix~A leaves four branch triples that can satisfy the degenerate matching condition. Their candidate sideband separations follow from a cubic equation in $\delta^2/4$, but the physical roots must be selected with the original unsquared equation because the algebraic reduction does not preserve the branch labels by itself.}

\rev{Direct pulse propagation confirms the three interbranch channels and also shows how the dynamics becomes progressively richer beyond the primary resonance. Configuration (II) produces a secondary fifth wave after the principal quartet has developed. More strikingly, within the explored parameter range the same-branch configuration (IV) is localized near the $\mathcal{CPT}$ threshold, where a single pump can satisfy two distinct sideband resonances and the ensuing evolution develops a multifrequency cascade. Away from this near-degenerate regime, the biorthogonal three-wave model captures the initial energy exchange and provides a transparent interpretation in terms of frequency-dependent nonlinear overlaps. Close to the exceptional point, however, the dual basis becomes ill-conditioned and additional frequencies grow early, so the failure of the few-mode reduction is itself a signature of the changing spectral geometry. The efficiency maps further distinguish a broad response for configuration (I) from a more selective response for configuration (II). Taken together, these results show that first- and second-order coupling dispersion play different but coupled dynamical roles: the first-order term introduces spectral asymmetry and modifies walk-off, while the second-order term controls the branch curvature, the extent of the unbroken-$\mathcal{CPT}$ domain, and the accessibility of near-threshold intrabranch resonances. More generally, the system provides an example of how a non-Hermitian spectral degeneracy can reorganize nonlinear resonances and delimit the regime in which a low-dimensional modal description remains reliable.}

\section*{Acknowledgments}
This research is funded by the Vietnam National Foundation for Science and Technology
Development (NAFOSTED) under grant number 103.01-2021.152.

\section*{Disclosures}
The authors declare that they have no known competing financial interests or personal relationships that could have appeared to influence the work reported in this paper.

\appendix

\section{Exclusion of non-phase-matched degenerate configurations}

Let $t=\delta/2>0$ and write the degenerate matching condition as
\begin{equation}
 2t^2+2s_1E_0-s_3E_+-s_4E_-=0,
 \label{eq:appendix-match}
\end{equation}
where
\begin{equation*}
 E_0=\varepsilon(\omega_p),\qquad
 E_+=\varepsilon(\omega_p+t),\qquad
 E_-=\varepsilon(\omega_p-t).
\end{equation*}
\rev{The purpose of this appendix is to justify the branch classification used in the main text without relying on numerical inspection of Eq.~\eqref{eq:degenerate-phase-matching}. The exclusion is conditional on the normalized weak-dispersion domain of the main analysis. After setting $\kappa_0=1$, we assume}
\begin{equation*}
\begin{aligned}
 0&<\kappa_2<1, & \kappa_1^2&<4\kappa_2,\\
 0&\leq\gamma\leq m, &
 m&\equiv\gamma_{\mathcal{CPT}}
 =1-\frac{\kappa_1^2}{4\kappa_2}>0.
\end{aligned}
\end{equation*}
\rev{The numerical examples additionally use $0<\kappa_1<1$, which expresses the weak-dispersion ordering adopted for the simulations but is not required by the proof below. The only bound on $\kappa_1$ needed analytically is the positivity condition $\kappa_1^2<4\kappa_2$, which ensures $m>0$. Introducing}
\begin{equation*}
 \omega_*=-\frac{\kappa_1}{2\kappa_2},
 \qquad
 \widehat K(\omega)=m+\kappa_2(\omega-\omega_*)^2,
\end{equation*}
shows that $\varepsilon(\omega)$ is even about $\omega_*$ and increases with $|\omega-\omega_*|$.

\subsection{Configuration \texorpdfstring{$(+,-,-)$}{(+,-,-)}}

For $(s_1,s_3,s_4)=(+,-,-)$, Eq.~\eqref{eq:appendix-match} becomes
\begin{equation*}
 2t^2+2E_0+E_++E_-=0,
\end{equation*}
which is impossible because every term is nonnegative and $t>0$.

\subsection{Configurations \texorpdfstring{$(+,-,+)$}{(+,-,+)} and \texorpdfstring{$(+,+,-)$}{(+,+,-)}}

For $(+,-,+)$,
\begin{equation}
 E_--E_+=2E_0+2t^2>0.
 \label{eq:mixed-exclusion-target}
\end{equation}
If $\omega_p\geq\omega_*$, monotonicity about $\omega_*$ gives $E_+\geq E_-$, contradicting Eq.~\eqref{eq:mixed-exclusion-target}. It remains to take $\omega_p<\omega_*$ and define $y=\omega_*-\omega_p>0$ and
\begin{equation*}
 f_\gamma(r)=\sqrt{(m+\kappa_2r^2)^2-\gamma^2},\qquad r\geq0.
\end{equation*}
Then Eq.~\eqref{eq:mixed-exclusion-target} is equivalent to
\begin{equation}
 D_\gamma(y,t)=f_\gamma(y+t)-f_\gamma(|y-t|)-2f_\gamma(y)-2t^2=0.
 \label{eq:Dgamma}
\end{equation}
For $0\leq\gamma<m$, all denominators below are positive and
\begin{equation*}
 \frac{\partial f_\gamma(r)}{\partial\gamma}=-\frac{\gamma}{f_\gamma(r)}.
\end{equation*}
Since $f_\gamma(y+t)\geq f_\gamma(|y-t|)$, it follows that $\partial_\gamma D_\gamma\geq0$. Hence $D_\gamma\leq D_\eta$ whenever $\gamma\leq\eta<m$. Taking the continuous limit $\eta\to m^-$ gives $D_\gamma\leq D_m$, including the threshold case where one of the arguments may vanish.

At $\gamma=m$,
\begin{equation*}
 f_m(r)=\kappa_2G_c(r),\qquad
 G_c(r)=r\sqrt{r^2+c^2},\qquad
 c^2=\frac{2m}{\kappa_2}.
\end{equation*}
Therefore
\begin{equation}
 D_m=\kappa_2\left[G_c(y+t)-G_c(|y-t|)-2G_c(y)\right]-2t^2.
 \label{eq:Dm}
\end{equation}
It remains to prove the bound
\begin{equation}
 G_c(y+t)-G_c(|y-t|)-2G_c(y)\leq2t^2.
 \label{eq:Gc-bound}
\end{equation}
For $c=0$, the left-hand side is $4yt-2y^2\leq2t^2$. For $c>0$, differentiate the left-hand side with respect to $c$. Apart from the positive factor $c$, the derivative is
\begin{equation*}
 h_c(y+t)-h_c(|y-t|)-2h_c(y),
 \qquad
 h_c(r)=\frac{r}{\sqrt{r^2+c^2}}.
\end{equation*}
The function $h_c$ is increasing and concave on $r\geq0$, with $h_c(0)=0$; consequently it is subadditive. If $t\leq y$, then
$h_c(y+t)\leq h_c(y)+h_c(t)\leq2h_c(y)$, and subtracting $h_c(y-t)$ makes the expression no larger. If $t\geq y$, concavity implies that an increment over an interval of length $2y$ decreases with the starting point, so
$h_c(t+y)-h_c(t-y)\leq h_c(2y)-h_c(0)\leq2h_c(y)$. The derivative is therefore nonpositive, and the expression is bounded above by its value at $c=0$, which proves Eq.~\eqref{eq:Gc-bound}. Combining Eqs.~\eqref{eq:Dm} and \eqref{eq:Gc-bound} yields
\begin{equation*}
 D_\gamma\leq D_m\leq-2(1-\kappa_2)t^2<0.
\end{equation*}
This contradicts Eq.~\eqref{eq:Dgamma}; hence $(+,-,+)$ has no nonzero solution.

The other mixed configuration is removed by reflection about $\omega_*$, without extending the convention $t>0$ to negative detuning. Suppose that $(+,+,-)$ were phase matched at $(\omega_p,t)$ and define the reflected pump frequency $\widetilde\omega_p=2\omega_*-\omega_p$. The symmetry $\varepsilon(2\omega_*-\omega)=\varepsilon(\omega)$ gives
\begin{equation*}
 \widetilde E_0=E_0,\qquad
 \widetilde E_+=E_-,\qquad
 \widetilde E_-=E_+.
\end{equation*}
The $(+,+,-)$ condition $E_+-E_-=2E_0+2t^2$ would then become
$\widetilde E_- -\widetilde E_+=2\widetilde E_0+2t^2$, which is exactly the already excluded $(+,-,+)$ condition at $\widetilde\omega_p$. Therefore $(+,+,-)$ has no nonzero solution either.

\subsection{Configuration \texorpdfstring{$(-,-,-)$}{(-,-,-)}}

For $(-,-,-)$, Eq.~\eqref{eq:appendix-match} requires
\begin{equation}
 E_++E_--2E_0=-2t^2<0.
 \label{eq:lower-branch-contradiction}
\end{equation}
For $\gamma<m$, direct differentiation gives
\begin{equation}
 \varepsilon''(\omega)=
 \frac{2\kappa_2\left[\widehat K^3-3\widehat K\gamma^2+2m\gamma^2\right]}
 {\left(\widehat K^2-\gamma^2\right)^{3/2}}\geq0.
 \label{eq:epsilon-convexity}
\end{equation}
Indeed, the bracket is an increasing function of $\widehat K\geq m\geq\gamma$ and at $\widehat K=m$ equals $m(m^2-\gamma^2)\geq0$. Thus $\varepsilon$ is convex for every $\gamma<m$. At $\gamma=m$, $\varepsilon_m$ is the finite pointwise limit of these convex functions and is therefore convex as well; equivalently,
\begin{equation*}
 \varepsilon_m(\omega)=|\omega-\omega_*|
 \sqrt{2m\kappa_2+\kappa_2^2(\omega-\omega_*)^2}
\end{equation*}
has an explicit convex cusp at $\omega_*$. Midpoint convexity yields
\begin{equation*}
 E_++E_-\geq2E_0,
\end{equation*}
contradicting Eq.~\eqref{eq:lower-branch-contradiction}. Hence $(-,-,-)$ is excluded.

Thus, within the stated parameter domain, the branch triples left for the main-text analysis are
\begin{equation*}
 (-,+,+),\qquad(-,-,+),\qquad(-,+,-),\qquad(+,+,+).
\end{equation*}

\end{document}